\documentclass[a4paper,useAMS,usenatbib]{mn2e}

\usepackage[total={17.8cm,24.0cm},centering]{geometry}
\usepackage[flushleft]{threeparttable}
\usepackage{graphicx}
\usepackage{float}
\usepackage{times}
\usepackage{color}
\usepackage{subcaption}
\usepackage{url}
\usepackage{arydshln}

\def\oiii{[O\,{\sc iii}]}

\def\hbeta{\hbox{H$\beta$}}
\def\ni{\hbox{[N\,{\sc i}]}}

\def\mcyl{$m_{\rm cyl}$}

\newcommand{\sauron}{{\texttt {SAURON}}}
\newcommand{\Xsauron}{{\texttt {XSAURON}}}

\title[Cylindrical rotation in B/P bulges]
{Establishing the level of cylindrical rotation in Boxy/Peanut bulges}

\author[Molaeinezhad et al.]
{A. Molaeinezhad$^1$\thanks{Email: molaei@ipm.ir},
J. Falc\'on-Barroso$^{2,3}$,
I. Mart\'inez-Valpuesta$^{2,3}$,
H.G. Khosroshahi$^{1}$,
\newauthor
M. Balcells$^{2,4}$, and
R.F. Peletier$^{5}$\\
$^{1}$School of Astronomy, Institute for Research in Fundamental Sciences (IPM), PO Box 19395-5746 Tehran, Iran \\
$^{2}$Instituto de Astrof\'isica de Canarias, E-38200, La Laguna, Spain\\
$^{3}$Depto. Astrof\'isica, Universidad de La Laguna (ULL), E-38206 La Laguna, Tenerife, Spain \\
$^{4}$Isaac Newton Group of Telescopes, Apartado 321, 38700 Santa Cruz de La Palma, Canary Islands, Spain \\
$^{5}$Kapteyn Astronomical Institute, University of Groningen, Postbus 800, 9700 AV Groningen, the Netherlands}

\begin{document}
 \date{Accepted 2015 November 16. Received 2015 November 05; in original form 2015 July 07}
\pagerange{\pageref{firstpage}--\pageref{lastpage}} \pubyear{2015}

\maketitle

\label{firstpage}

\begin{abstract}
We present \sauron\ integral-field observations of a sample of 12 mid to high-inclination
disk galaxies, to unveil hidden bars on the basis of their kinematics, i.e., the correlation between velocity and $h_3$ profiles, and to establish their degree of cylindrical rotation.
For the latter, we introduce a method to quantify cylindrical rotation that is robust
against inner disk components. We confirm high-levels of cylindrical rotation in boxy/peanut bulges,
 but also observe this feature in a few galaxies
with rounder bulges. We suggest that these are also barred galaxies with end-on
orientations. Re-analysing published data for our own Galaxy using this new
method, we determine that the Milky Way bulge is cylindrically rotating at the
same level as the strongest barred galaxy in our sample. Finally, we use
self-consistent three-dimensional $N$--\,body simulations of bar-unstable disks
to study the dependence of cylindrical rotation on the bar's
orientation and host galaxy inclination.


\end{abstract}


\section{Introduction}
\label{sec:intro}

Barred galaxies represent a considerable fraction of the entire disk galaxy
population \citep[e.g.][]{eskr2000,knapen2000,whyt2002,grosb2004,marin2007}. In
edge-on or highly inclined systems, bars are most easily recognised by boxy and peanut-shapes, and sometimes X-shape morphology formed by the stellar material above the disk plane
\citep[e.g.][]{kuij1995,bure1999}. Bars in this configuration are usually termed
Boxy/Peanut bulges (hereafter BP bulges).

$N$--\,body simulations of bar-unstable disks indicate that BP bulges could form out of disc material, via the
 vertical buckling instability of the bar \citep[e.g.][]{raha1991,mart2004,mart2006}. However, including gas in the simulations may suppress buckling and as a result, the peanut can form without a vertical buckling phase  \citep[e.g.][]{bere1998,bere2007,vill2010,deba2006,wozn2009}.  \citep[For a general review of this subject, see][]{atha2015}.

Finding observational evidence to establish a link between bars and
 BP shape is sometimes complicated. 
The vertically thickened regions of bars are not observable photometrically
if the galaxy is not inclined enough \citep[e.g.][]{atha2006,erwi2013}, while the 
presence of a bar is generally apparent, on the basis of photometry alone, if 
the galaxy is viewed in a more face-on orientation.
 In fact, the best viewing angles, giving information on both the BP and bar, are 
 intermediate inclination, but close to edge-on \citep[see][and the references therein]{atha2015}.
BP bulges are best seen when the bar is at an intermediate angle or perpendicular to the
line-of-sight. Bars oriented exactly parallel to the line-of-sight appear almost
spherical and thus difficult to identify \citep{bure2004, atha2015}.

BP bulges specifically and bars in general produce distinct kinematic features
that can be easily detected in galaxies. \citet{bure2005} established, using
$N$--\,body simulations, a number of kinematic properties in bars of different
strength and orientations in highly-inclined systems. They are most noticeable
in the major-axis rotation and velocity dispersion profiles (e.g.
``double-hump'' rotation curves, velocity dispersion profiles with a plateau at
moderate radii), but they also leave more subtle signatures in the higher order
moments of the Gauss-Hermite series (e.g. a positive correlation between the
velocity and the $h_3$ over the length of the bar). These features have been
confirmed in several observational studies \citep[e.g.][]{1981ApJ...247..473P,
1983ApJ...275..529K, 1997A&AS..124...61B, 2001A&A...368...52E,
2003A&A...409..459M, 2009A&A...495..775P, will2011}. While being a powerful
indicator for the presence of bars, the $V-h_3$ correlation has hardly been
used for this purpose in observational studies
\citep[e.g.][]{chun2004}. All these diagnostics have been recently
expanded \citep{2015MNRAS.450.2514I} to be applied to the wealth of data from
existing or upcoming integral-field spectroscopic surveys (e.g. ATLAS3D,
\citealt{2011MNRAS.413..813C}; CALIFA, \citealt{2012A&A...538A...8S}; SaMI,
\citealt{2012MNRAS.421..872C}; MaNGA, \citealt{2015ApJ...798....7B}).

 \begin{figure*}
 	\captionsetup[subfigure]{labelformat=empty}
 	\centering
 	\begin{subfigure}{1.0\textwidth}
 		\centering
 		\includegraphics[width=1.0\textwidth]{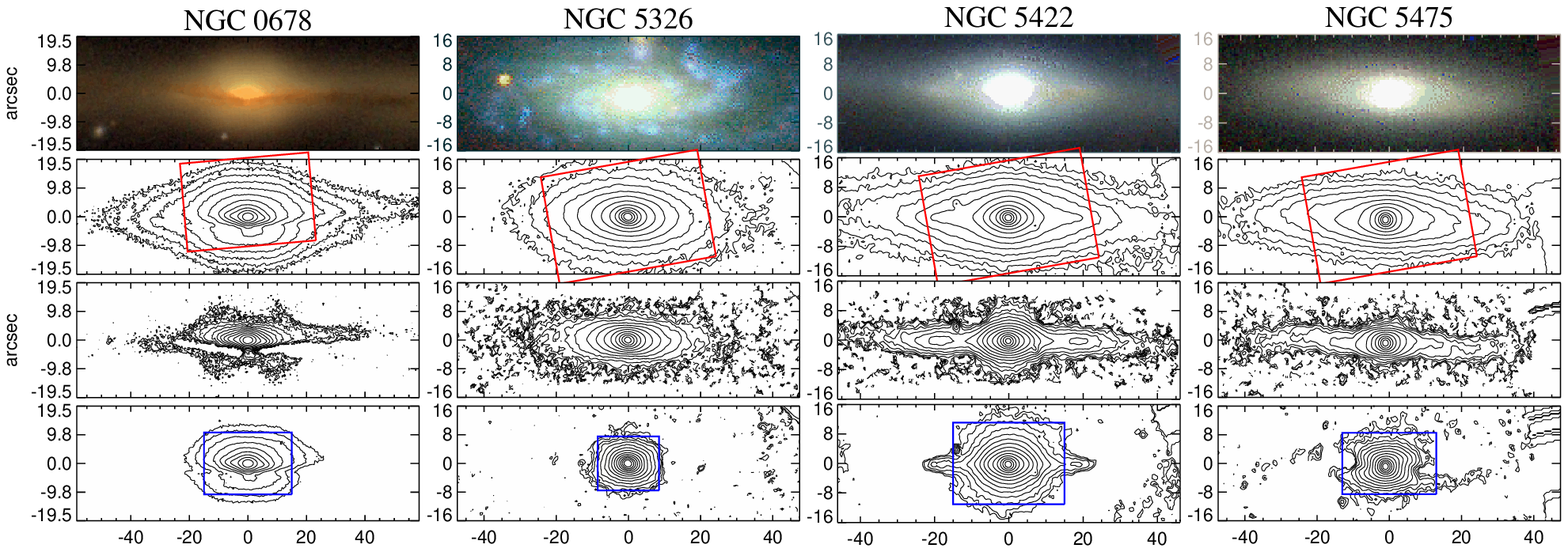}
 	\end{subfigure}
 	
 	\begin{subfigure}{1.0\textwidth}
 		\centering
 		\includegraphics[width=1.0\textwidth]{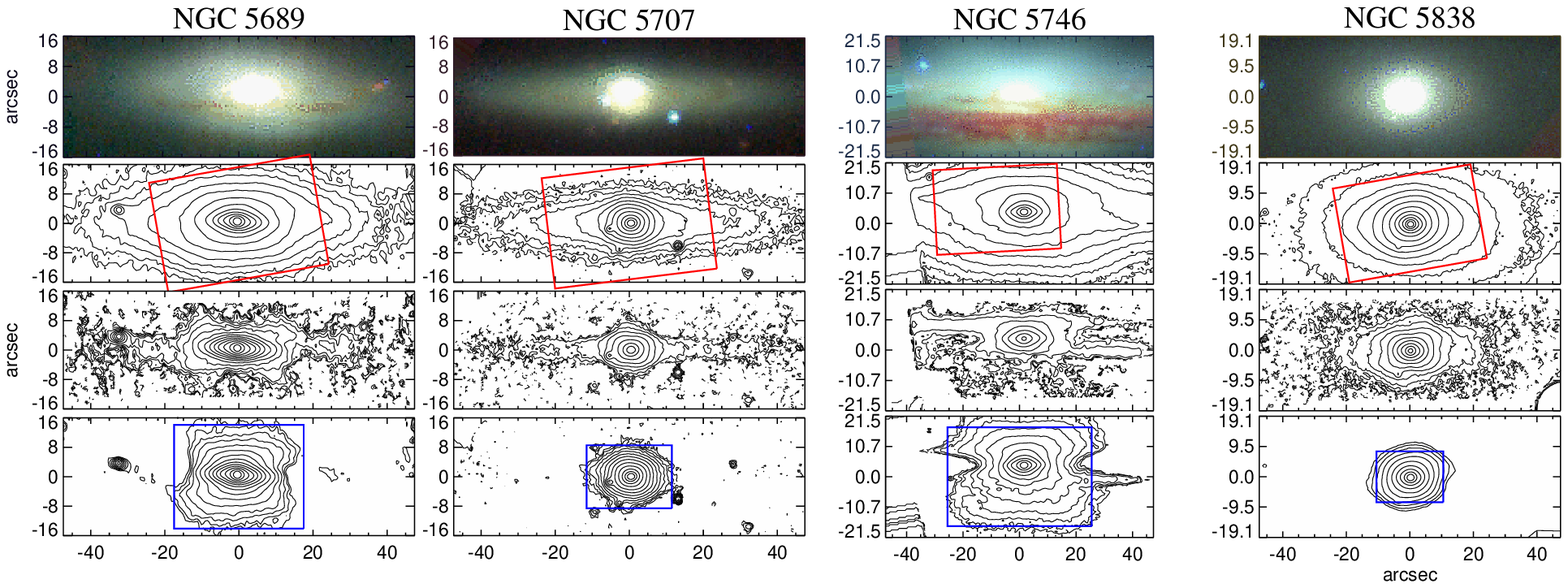}
 	\end{subfigure}
 	
 	 	\begin{subfigure}{1.0\textwidth}
 	 		\centering
 	 		\includegraphics[width=1.0\textwidth]{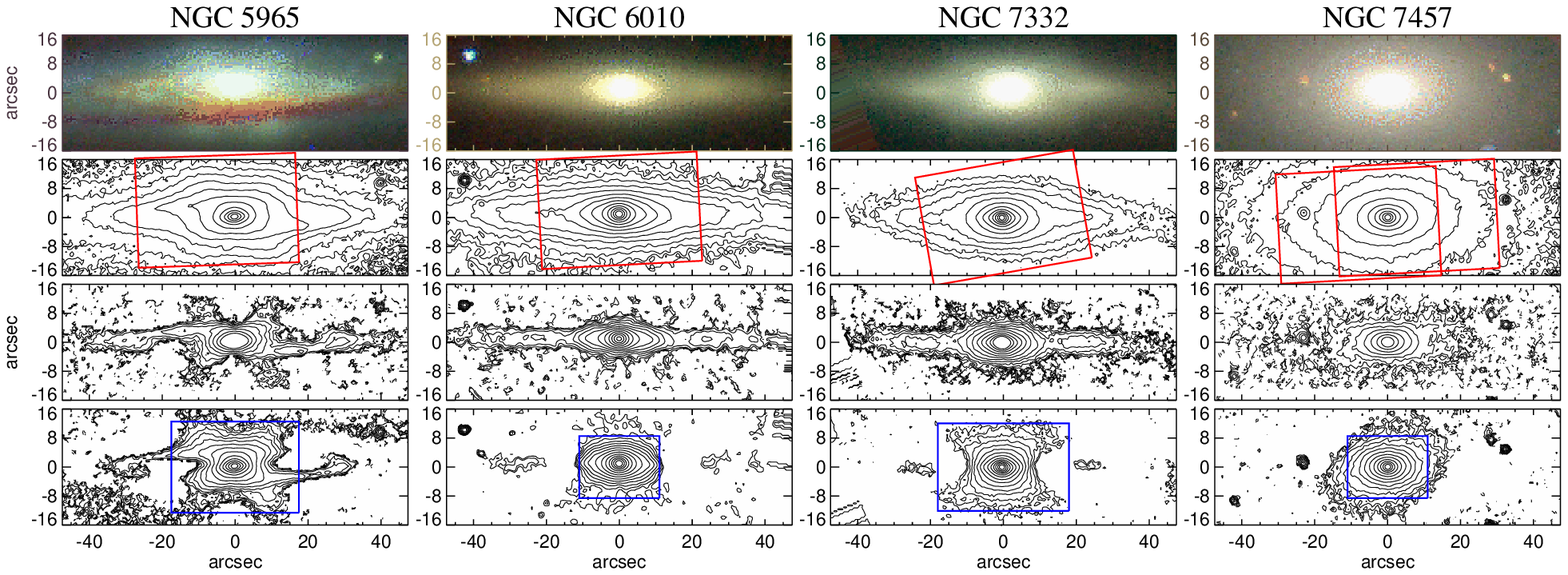}
 	 	\end{subfigure}
 	\setcounter{figure}{1}
 	\centering
 	\centering
 	\caption{$U$-$R$-$K$ composite images, Isophotes and unsharp masks for all 12 galaxies in our sample. For each galaxy, First panel: The $U$-$R$-$K$ composite image, from \citet{pele1997} . $U$-- and $R$-band images were obtained with the Prime Focus camera of the Isaac Newton telescope and $K$ -band images were taken at the UK infrared telescope (UKIRT), using the IRCAM3 Camera. For NGC\,0678 we have used Sloan Digital Sky survey DR10 data.
 		Second panel: Isophotes of the $K$-band images. The \sauron\ final pointing is shown on top of this panel. Third panel: Unsharp masked $K$-band image. The unsharp masking technique allows us to highlight the structure of the galaxy in more details, for instance, the X-shape bulge is clearly visible for 6 galaxies in our sample (NGC\,0678, NGC\,5422, NGC\,5689, NGC\,5746, NGC\,5965 and NGC\,7332). Fourth Panel: Isophotes of the $K$-band image, after subtracting an exponential disk from the observed image. We have used the disk-subtracted images together with the unsharp masked galaxy images (third panel) to have a rough estimation of bulge size in the $(x,z)$ space. This analysis bulge window is shown as a blue rectangle on top this panel. }
 	\label{fig:allphot}
 \end{figure*}


Cylindrical rotation \citep{korm1982} is another major kinematic characteristic
feature of BP bulges \cite[e.g.][]{korm1982,bure1999,falc2006}. In such systems,
the mean stellar rotation speed is almost constant with height above the disk
plane. This feature is well described by the formation and evolution scenarios
of BP bulges through the vertical buckling instability of the bar
\citep{comb1981,mart2004} and, subsequently, many $N$--\,body simulations of barred
galaxies \citep[e.g.][]{comb1990,sell1993,atha2002,bure2005,saha2012,mart2013,sahab2013} present
such feature in the velocity maps. \citet{will2011} studied the stellar
kinematics and populations of BP bulges in a relatively small sample of five
edge-on galaxies using long slit spectroscopy. Their main conclusion is that the
degree of cylindrical rotation varies significantly among BP bulges, as do the
properties of their stellar populations. That study was the first to challenge
the simple picture of BP bulges and claimed that they do not form
a homogeneous class of objects. \citet{saha2012} investigated the interaction of
a bar and a low-mass classical bulge via a high spatial resolution $N$--\,body
simulation of a galaxy consisting of a live disc, a bulge and dark matter halo.
They followed the evolution of the kinematics of the small classical bulge. In
their study, they show that the classical bulge could acquire cylindrical
rotation by exchanging angular momentum with the surrounding bar.
 Later on, \citet{saha2013} introduced a criterion to quantify deviations from pure
cylindrical rotation in edge-on disk galaxies and use this quantity to show that
the level of cylindrical rotation in BP bulges is time-dependent.

This paper presents the first results of our analysis of the properties of BP bulges in nearby galaxies. 
Here we will focus on the stellar kinematics of a sample of 12 intermediate inclined disk galaxies
($i > 60^{\circ}$) to unveil the presence of bars and to quantify the level of
cylindrical rotation in these systems. In future papers we will connect these
features with the stellar population properties of these galaxies. The paper is
structured as follows. In Section~\ref{sec:sample} we introduce our sample and
the photometric and spectroscopic datasets used for this work.
Section~\ref{sec:corrprofs} we study the signature of the bar in our 2D
kinematics maps and methods to unveil the hidden or end-on bars based on
$N$--\,body simulations results. Section~\ref{sec:cylrot} is devoted to discuss the
signature of cylindrical rotation and methods to describe it in BP bulges. In
Section~\ref{sec:mw_cyl}, we focus on the determination of the cylindrical
rotation in the Milky Way(MW) bulge. Section~\ref{sec:discussion} discusses the possible
correlation between bars, BP bulges and level of cylindrical rotation. We
summarize our results and draw our conclusions in Section~\ref{sec:conclusions}.

\section{Sample \& IFU observations}
\label{sec:sample}

\subsection{Sample Selection \& Properties}
\label{sec:sam_pro}
Our sample for this study comprises 12 intermediate inclined galaxies
($i>60^{\circ}$) selected from the well-characterised sample of \citet{balc1994}, of inclined early-type spiral galaxies and S0's.\footnote{The optical and near-infrared (NIR) surface brightness and colour profiles of galaxies in the original sample of  are available in the web page \url{http://www.astro.rug.nl/~peletier/www/newast.html}}
This preferred orientation allows us to use the relatively little obscured side of the bulge.
The sample was complemented with NGC\,0678, that followed the same selection criteria and had similar
\sauron\ integral-field spectroscopic observations.

The \sauron\ data for the galaxies in our sample have not been published before, apart from NGC\,5475 and NGC\,5689 \citep{falc2006} and NGC\,7332 \citep{falc2004}. For these 3 galaxies, the data published here is the same as was presented before. However, in these papers, the issue of cylindrical rotation was not addressed. The other 9 galaxies, although observed with SAURON, are not part of the \sauron\ sample, nor of the sample of late-type spirals of \cite{gand2006}.
The sample was observed independently, as part of a project to study the vertical kinematics and population structure of bulges. The observed field of view and spatial resolution is well matched to the sizes of the bulges. The $S/N$ of the data is considerably higher than that of the ATLAS3D survey \citep{capp2011}, on which we will elaborate in the second paper of this series. Another advantage of this sample is that the extinction properties have been well-studied \citep{pele1999}. We present all the kinematic maps and new diagnostics developed in this paper in Appendix~\ref{app:cylrot}.

These objects have been the subject of
several studies on galactic bulges by the original authors
\cite[e.g.][]{pele1996,pele1997, pele1999, falc2002,
2003ApJ...582L..79B,balc2003, falc2003a, falc2003b,
2007ApJ...665.1084B, 2007ApJ...665.1104B} and have extensive ground-based and
{\it Hubble Space Telescope} optical observations as well as NIR photometry.

It is worth noting that in this study, we do not label our bulges as classical or pseudobulges or bars, trying to make the analysis independent of the photometric appearance of the bulges. Through this study, the 'bulge' of a galaxy is referred to the excess of light that is above the main disk. 
On the other hand, it is our aim to do a study that is as independent as possible from photometric studies in the past, as obvioulsy by performing a decomposition into various types of bulges, the results will be severely biased.
This way of defining the bulge might bias against certain type of 
bulges like as pseudo bulges. In \S\ref{sec:corrprofs} we will briefly discuss the masking
effect of a classical bulge on the kinematics signatures of barred systems.

\subsection{Morphology Analysis}
The overall shape of the non-axisymmetric component in highly inclined galaxies
is not directly visible in the images, but becoming clear in unsharp masking
 or residual images in decompositions. Figure~\ref{fig:allphot} (second and third panels) shows the isophotes 
 and unsharp masks of the $K-$band images for all 12 galaxies in our sample.
Our $K-$band images were obtained with the UK infrared 
telescope (UKIRT), using the IRCAM3 Camera \citep{pele1997}. 
For NGC\,0678 we have used $i-$band Sloan Digital Sky survey DR10 data \citep{2014ApJS..211...17A}.
 The images after unsharp masking (median filtering) show the
high frequency features of the images and an X-shaped structure is clearly visible for 
6 galaxies in our sample (NGC\,0678, NGC\,5422, NGC\,5689, NGC\,5746, NGC\,5965 
and NGC\,7332). The lower panels of Figure~\ref{fig:allphot} presents the isophotes of the $K-$band 
image, after subtracting an exponential disk from the observed image. For this 
purpose, we used GALFIT \citep{peng2002} to find the best exponential disk model 
while the central regions (0.5 kpc) of galaxies are masked to avoid any possible influence 
of bright central components. As our goal was to model the outer disks, we found such 
masking approach more useful than putting any constraints on the fitting parameters, while 
the radius of this masked region is large enough to avoid large bulges from affecting the disk fit.
 The fitting process has been repeated using different initial parameters to obtain the best 
 disk model. In the fourth panel of Figure~\ref{fig:allphot}, for each galaxy, we show the residual 
 image, obtained after subtracting the disk from the original image.

 It is obvious that ignoring the bar component from the fitting process will introduce a bias in the fitted parameters of the bulge and the disk. As also stated in \citet{gado2008} both bulge and disk components will try to accommodate the light from the ignored bar. The resulting disks will have a steeper luminosity profile,  and a stronger effect will be seen in the bulge which will acquire a larger effective radius and luminosity fractions. This works in favor of our goal which is not to miss any possible bulge contribution. Again we stress that the purpose of this exercise is to obtain a rough estimation for the bulge analysis window by modeling out the exponential disk. We use the term "photometric bulge" to indicate that we do not discriminate between different types of bulges in this study.

Evaluation of the photometric bulge analysis radius in $x$ (or $z$) direction has been performed by
stacking the residual images (single exponential disk subtracted), along the minor (or major) axis and evaluating
the radius, at which 90\% of the galaxy light is concentrated within that
radius. The 90\% was preferred to lower typical values such as 75 or 50\% just to
make sure that the whole bulge area is covered during our analysis.
 However, selecting different thresholds (50, 75 or 90\%) for the bulge radius
 does not impose any major difference, as the brightness profile of the bulge 
 is much steeper than that of the disk. 
 Comparing the third and fourth panel of Figure~\ref{fig:allphot} for each galaxies
 demonstrates good agreement between this approach to evaluate the bulge
  analysis window and that one could obtain using unsharp masked image.
  
  Clearly, the radial extent parameters $x_{B}$ and $z_{B}$ shrink if a luminous spheroidal component exists in the center of a box-peanut light distribution. This is acceptable to us, given that we are after the kinematics of everything, excluding the exponential disk. Any possible bias in our results due to our proposed method to evaluate the bulge analysis windows will be discussed later.

In our study, we are interested in determining the level of cylindrical
rotation of these bulges and thus it is important how many of them are BP
bulges and how many are largely spherical. It is widely accepted that fourth
order Fourier coefficients ($c_4$) of an ellipse-fit extraction can be used to
measure the level of boxiness or diskiness of different structural components in
galaxies. This method has been used by many authors to quantify the degree of
boxiness in bulges \citep[e.g.][]{comb1990,shaw1993,lutt1996,merr1999}. We
attempted to carry out this type of analysis in our sample, but we were not
successful to obtain robust results. Difficulties arose due to the amount of
dust present in many of these galaxies. Being close to edge-on, dust in the main
disk is very prominent and thus complicates the extraction of reliable
ellipse-fit profiles. The particular
orientation of our galaxies, also causes that different structures with very
distinct $c_4$ signal (e.g. boxy bulges and inner disks) overlap in projection
and produce an ambiguous $c_4$ value. 
Such inconsistencies between the visual
inspections and the results of isophotal analysis have been discussed in the
literature \citep[e.g][]{lutt2000}. As they argued, a major disadvantage of using
 isophotal analysis is not only the presence of dust, but the influence of the masked stars in the foreground, presence of non-axisymmetric structure like flat bars, thick disk, inner disks and the extreme nature of BP isophotal distortions for the ellipse fitting.

In order to overcome this issue, and similar to \citet{lutt1996}, we 
evaluate the level of boxiness of bulges in our sample by visually inspecting
residual images after subtracting an exponential disk model to the galaxies. 
As we mentioned earlier, the result of this exercise on our $K-$band images of our sample is presented in
Figure~\ref{fig:allphot}.

Our analysis of the residual images in addition to the unsharp masked images 
suggest that 6 galaxies host prominent BP bulges (NGC\,0678, NGC\,5422, NGC\,5689, 
NGC\,5746, NGC\,5965 and NGC\,7332), while the rest harbour relatively spherical bulges 
or there is no clear sign of X-shape structure in the related unsharp masked images. 
Our morphological classification of these
bulges is consistent with the previous studies \citep{lutt2000,chun2004,will2011}.
The basic photometric properties of all 12 galaxies in our sample are summarised in Table \ref{tab:sample}.

\begin{table*}
	\caption {Basic properties of our sample of galaxies}
	\label{tab:sample}
	\begin{center}
		\begin{tabular}{llllllllll}
			\hline
			Galaxy           & $V_{LG}$        & Scale       & B/D  &   Inclniation &    Type   &    Type         & PA&     $x_{B}$    & $z_{B}$     \\
			~                   & ($kms^{-1}$) & ($kpc/''$)  &   ~    &  ($deg$)          &   ($RC3$)   &($Buta+2015$) & ($deg$)                    & ($arcsec$)   & ($arcsec$)    \\
			(1)                 &   (2)               &  (3)            & (4)    &(5)                &   (6)        &   (7)            &  (8)    & (9)            & (10)         \\
			\hline     
			NGC\,0678    &   3015 & 0.19   &  -       &  82             & $SB(s)b$ $sp$     & $SAB(s,nd)a$ $sp$                    & 9      & 15.0       &  10.5   \\
			NGC\,5326    &   2596  & 0.17  & 0.73  &  65             & $SAa?$                & $E(d)5/S0^{–}$ $sp$                  & -138   & 8.5        &  7.5    \\
			NGC\,5422    &   1977  & 0.13  & 1.28  &  90(90)       & $S0$ $sp$           & $SAB_{ax}0^{o}$ $sp/E(d)8$           & 64     & 15.0       &  11.0   \\
			NGC\,5475    &   1815  & 0.12  & 0.14  &  78(79)       & $Sa?$ $sp$         & -                                    & -101   & 13.0       &  8.5    \\
			NGC\,5689    &   2295  & 0.15  & 0.89  &  81             & $SB(s)0/a$           & $(R^{'}L)SAB:(r^{'}l,nd)0^{+}$ $sp$  & -5     & 17.5       &  14.0   \\
			NGC\,5707    &   2358  & 0.15  & 0.42  &  80             & $Sab?$ $sp$	    & $SA0^{o}$ $sp$	                   & -59    & 11.5       &  8.5    \\
			NGC\,5746    &   1680  & 0.12  & 0.68  &  81             & $SAB(rs)b?$ $sp$  & $(R')SB_{x}(r,nd)0/a$ $sp$           & -99    & 25.5       &  17.5   \\
			NGC\,5838    &   1316  & 0.09  & 0.71  &  72             & $SA0^{-}$	           & -	                                   & 134    & 10.5       &  8.0    \\
			NGC\,5965    &   3603  & 0.23  & 0.53  &  80             & $Sb$                    & -                                    & -38    & 17.5       &  12.5   \\
			NGC\,6010    &   2036  & 0.14  & 0.27  &  84(90)       & $S0/a ?$ $sp$     &$ SA(l)0^{o}$ $sp$                     & 12     & 11.0       &  8.5    \\
			NGC\,7332    &   1481  & 0.08  & 0.41  &  81(84)       & $S0$ $pec$ $sp$ & -                                    & -114   & 18.0       &  12.0   \\
			NGC\,7457    &   1134  & 0.06  & 10.0  &  70(74)       & $SA0^{-}(rs)?$     & -                                    & 35     & 11.0       &  8.5     \\
			\hline
			\end{tabular}
			\end{center}
		\begin{flushleft}
	\small NOTES: (1) Galaxy name. (2) Recession velocity of each galaxy in $km/s$, corrected to the Local Group (Karachentsev and Makarov 1996). (3) Spatial scale in kpc/arcsec at the galaxy distance, assuming $H_{0}=70$ $km/s/Mpc$. (4) Bulge-to-Disk luminosity ratio from a bulge-disk decomposition in $R-$band \citep{pele1997}, following \citet{kent1984} method. (5) Galaxy inclination evaluated from disk ellipticity in $R-$band from \citet{pele1997}, corrected for finite disk thickness. Inclinations evaluated from the best fitting mass-follow-light JAM models \citep{capp2013} are also mentioned, if available. (6) Morphological type from the Third Reference Catalog of Bright Galaxies \citep[][hereafter RC3]{deva1991}. (7) Morphological classification by \citet{buta2015} from the Spitzer Survey of Stellar Structure in Galaxies ($S^{4}G$). (8) The position angle (N--E) of the dust free minor axis. (9) and (10) Analysis window of the bulge along the major and minor axis respectively. Refer to text for more details.
           \end{flushleft}
\end{table*}

\subsection{Integral-field observations and data reduction}

Spectroscopic observations were carried out between October 1999 and 2011 with
the \sauron\ integral-field spectrograph \citep{baco2001} attached to the 4.2-m William
Herschel Telescope (WHT) of the Observatorio del Roque de los Muchachos at La
Palma, Spain. We used the low spatial resolution mode of \sauron\, which gives
a $33\arcsec\times41\arcsec$ field-of-view (FoV), with a spatial sampling of
$0\farcs94\times0\farcs94$. This setup produces 1431 spectra per pointing over
the \sauron\ FoV. Additionally, a dedicated set of $146$ lenses provide
simultaneous sky spectra $1\farcm9$ away from the main field. The spectral
resolution delivered by the instrument is $\sim$4.2~\AA\ (FWHM) and covers the
narrow spectral range 4800-5380~\AA. This wavelength range includes a number of
important stellar absorption features (e.g. \hbeta, Fe5015, Mg$b$, Fe5270) and
also potential emission lines (\hbeta$\lambda4861$,
\oiii$\lambda\lambda$4959,5007, \ni$\lambda\lambda$5198,5200).

For each galaxy, typically four overlapping exposures of 1800\,s were typically
obtained. An offset of a few arcseconds was introduced between exposures to
avoid bad CCD regions. Single pointings were required to cover the bulge
dominated region. We followed the procedures described in \citep{baco2001} for
the data reduction of the data using the specifically designed \Xsauron\
software developed at CRAL. For each galaxy, the sky level was measured using
the dedicated sky lenses and subtracted from the target spectra. Arc lamp
exposures were taken before and after each target frame for wavelength
calibration. Tungsten lamp exposures were also taken at the beginning and end of
each night in order to build the mask necessary to extract the data from the CCD
frames. Flux standard stars were observed during each observing run for
calibration purposes. The individually extracted and flux calibrated datacubes
were finally merged by truncating the wavelength domain to a common range and
spatially resampling the spectra to a common squared grid. The dithering of
individual exposures enabled us to sample the merged datacube onto
$0\farcs8\times0\farcs8$ pixels.

In order to ensure the measurement of reliable stellar kinematics, we spatially
binned our final datacubes using the Voronoi 2D binning algorithm of
\citet{2003MNRAS.342..345C}, creating compact bins with a minimum
signal-to-noise ratio ($S/N$) of $\sim40$ per spectral resolution element. Most
spectra in the central regions, however, have $S/N$ in excess of $40$, and so
remain un-binned. The stellar kinematic maps used in this paper (velocity [$V$],
velocity dispersion [$\sigma$], $h_3$ and $h_4$) were obtained following the
same procedure outlined in \citet{falc2006} using the penalised pixel fitting
(pPXF) routine by \citet{2004PASP..116..138C} and the MILES stellar population
models as templates \citep{2010MNRAS.404.1639V,2011A&A...532A..95F}.

\begin{figure*}
\centering
\includegraphics[width=1.02\textwidth]{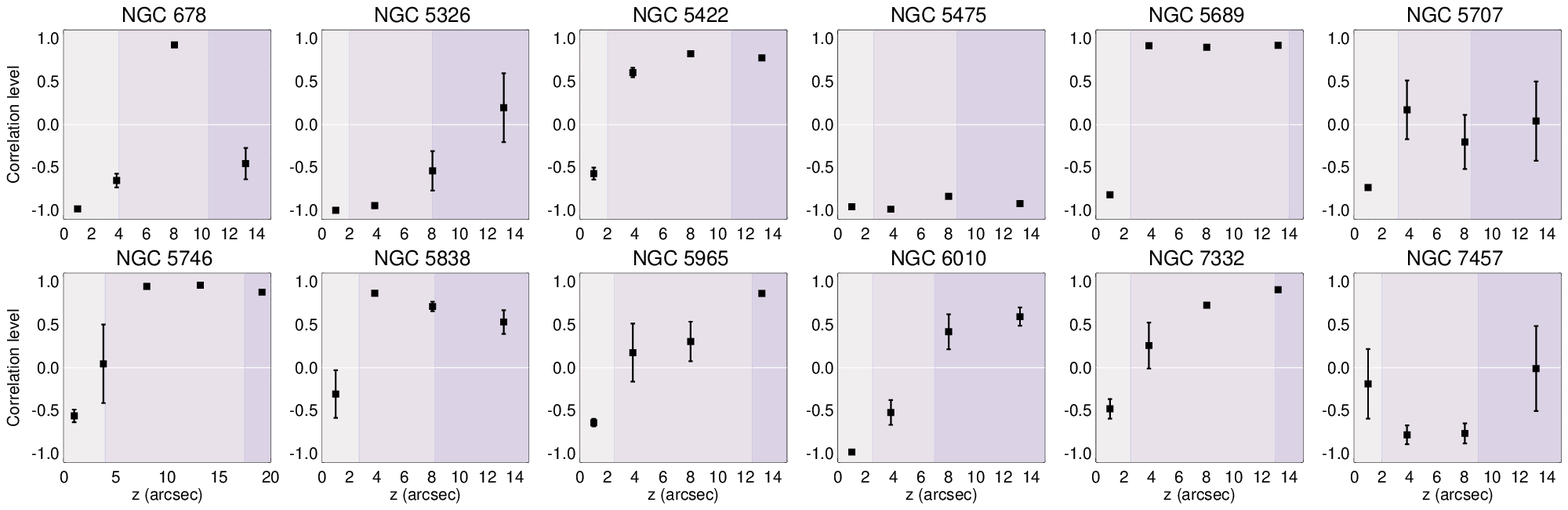}
\caption{The $V-h_3$ correlation profiles for all galaxies in our sample.
These profiles show the level of correlation of $V$ and $h_3$ within the bulge zone in pseudo-slits
parallel to the major axis of the galaxies at different heights ($z$) from the disk
plane. These measurements have been performed on the dust-free side of the
galaxies. Negative values indicate negative correlation, while values closer
to +1 mean stronger positive correlation. Different shaded regions mark the disk
plane, the bulge/bar dominated parts and the regions beyond $z_{B}$.}
\label{fig:corrprofs}
\end{figure*}

\section{Unveiling hidden bars with kinematics}
\label{sec:corrprofs}

$N$--\,body simulations of bar-unstable disks have established that BP bulges
are simply the thick part of bars extending above the galaxy disk. Our ability
to detect bars via photometry depends on the viewing angle of the bar and
galaxy inclination, as bars oriented parallel to the line of sight could
remain undetected. An alternative to the photometric observations is the use of
the stellar kinematics. As mentioned earlier, self-consistent, three-dimensional
$N$--\,body simulations of bar-unstable disks \citep[see][for more
details]{bure2005} suggest a number of the major-axis stellar kinematic features that can be used as
bar diagnostics: (1) a "double-hump'' rotation curve; (2) a broad central
velocity dispersion peak with a shoulder or plateau (and possibly a secondary
maximum) at moderate radii; and (3) an $h_3$ profile correlated with $V$ over
the projected bar length. They also showed that a local central velocity
dispersion minimum can also be present for strong bars seen approximately
end-on.

Following this work, \citet{chun2004} presented long-slit stellar kinematic profiles 
of 30 edge-on spiral galaxies, mostly with BP bulges. The
one-dimensional kinematic profiles of most galaxies in their sample were
consistent with those predicted from $N$--\,body simulations. Nevertheless, they
also pointed out a number of differences, which are most likely due to the
absence of a dissipative component in such simulations
\citep[see][]{frie1993,wozn2003}. They found that in 60\% of galaxies in their
sample $h_3$ is strongly anti-correlated with $V$ in the very central regions.
They related this feature to the presence of cold and dense decoupled central
stellar disks, and argued this inner disk formed out of gas accumulated by the
bar at the centre through inflow. Their results, however, confirmed that the
positive correlation of $V$ and $h_3$ can also be considered a reliable
indicator of bars.\looseness-2
 
  Recently, \citet{2015MNRAS.450.2514I} studied the imprints of boxy/peanut structures
   on the 2D line-of-sight kinematics of simulated disk galaxies. They recover the results 
   of \citet{bure2005} as well as their dependence on the strength of the
    boxy/peanut and its position angle and various projection effects. To this, they also
     add the peanut related signatures in the form of elongated wings of large $h_3$ values
      and X-shaped regions of deep $h_4$ minima, roughly in an area covering the peanut
       \citep[see][for more details]{2015MNRAS.450.2514I, atha2015}.

We have taken advantage of this diagnostic tools to unveil the presence of
hidden bars in our sample. In this section we will focus in particular on the
$V$ and $h_3$ correlation. We have used the correlation between $V$ and $h_3$ to
 unveil the presence of hidden bars in our sample. For this purpose, we have extracted
  profiles of these two parameters on pseudo-slits parallel to the major axis, at various
   heights from the major axis. These profiles are extracted only on the €œclean€
    (i.e., dust-free) side of our galaxies to avoid issues with dust extinction.
     Prominent cases with clear dust disks are NGC\,5746 and NGC\,5965.
      We have then determined the linear Pearson
correlation coefficient \citep[e.g.][]{093570275X} between $V$ and $h_3$ for
each pseudo-slit. The resulting correlation values are shown in
Figure~\ref{fig:corrprofs} for the entire sample. The error bar for each point
represents the standard error of the correlation coefficient \citep{ghos1966}.
In these plots, a value of -1 indicates a negative correlation and +1 means
positive correlation. Light purple shaded areas mark regions closer to the disk
plane, where signs of an anti-correlation due to the presence of inner disks are
expected. The intermediate regions mark the bulge/bar dominated regions where,
if present, we expect to measure a positive correlation between $V$ and $h_3$. 

The inspection of these profiles reveals that most (8 out of 12) galaxies in our
sample display positive correlation in the areas above the mid-plane. This
(positive) correlation is most visible in those systems with BP bulges.
 Our data therefore confirms previous claims that positive
 $V-h_3$ correlation is strongly associated with bars.
That several galaxies show positive $V-h_3$ correlation well outside the region of the photometric bulge is an indicator that photometric decomposition does not clearly separate regions with distinct bulge and disk dynamics. If regions of positive $V-h_3$ correlation are dynamically set by the action of bars, our result suggests that, at least in high-inclination views, such regions extend out beyond the photometric bulge and merge smoothly with the outer exponential disk.

Moreover, 10 of 12 galaxies in this sample, clearly exhibit a
strong anti-correlation ($<-0.5$) of $h_3$ and $V$ in the central regions,
similar to what is observed by \citep[e.g.][]{chun2004,falc2006} in barred
galaxies, and associated to rotating inner structures. In our sample, all galaxies with BP bulges show this feature in their central
regions.

As mentioned earlier, \citet{2015MNRAS.450.2514I} investigated the dependencies between 
the various projection effects and bar strength on the kinematic diagnostics
of barred system. They show that for stronger BPs, the kinematic maps change
considerably from the side-on to the end-on projection in both the amplitude
of the moments and their morphology and involve interesting features offset
 from the major axis. Specifically, the $V-h_3$ correlation
blobs become larger and extend farther out from the centre, while for 
moderate BPs the changes are considerably reduced and are essentially
in the magnitude of the moments.
Considering this issue, we can study a peculiar galaxy in this analysis: NGC\,5838.
Despite having a nearly spherical bulge, NGC\,5838 displays a high level of
correlation between $h_3$ and $V$ over the bulge region.
NGC\,5838 shows also properties akin of barred galaxies: double-hump
rotation curve or characteristic velocity dispersion profile. It is for this
reason that we argue that NGC\,5838 is most probably a barred galaxy seen close to
end-on (i.e. with the bar parallel to the line-of-sight). 

The second peculiar case in our sample is
NGC\,6010, which is interesting for various reasons. While the level of positive
correlation in the fairly round bulge region is low, it is the only case where
this correlation becomes very high beyond the bulge. In order to interpret the 
unusual behaviour of this galaxy, we refer to another interesting part of the study by 
\citet{2015MNRAS.450.2514I} which discusses the masking effects of  a classical 
bulge on the BP kinematic features. They found that the characteristic signatures 
of the bar are considerably weakened in the presence of a classical, spherically-symmetric 
component.

We suggest that NGC\,6010 is a barred galaxy with an spherically-symmetric 
central component, in which the vertical extent of the bar is beyond that of the classical 
bulge in the center. It may also address the issue that our method to define the bulge 
analysis window for this galaxy is biased against the presence of a classical bulge in the 
center. In the other words, the real vertical extent of the bar is further than we already evaluated.

%

\begin{figure*}
\centering
\includegraphics[width=1.03\textwidth]{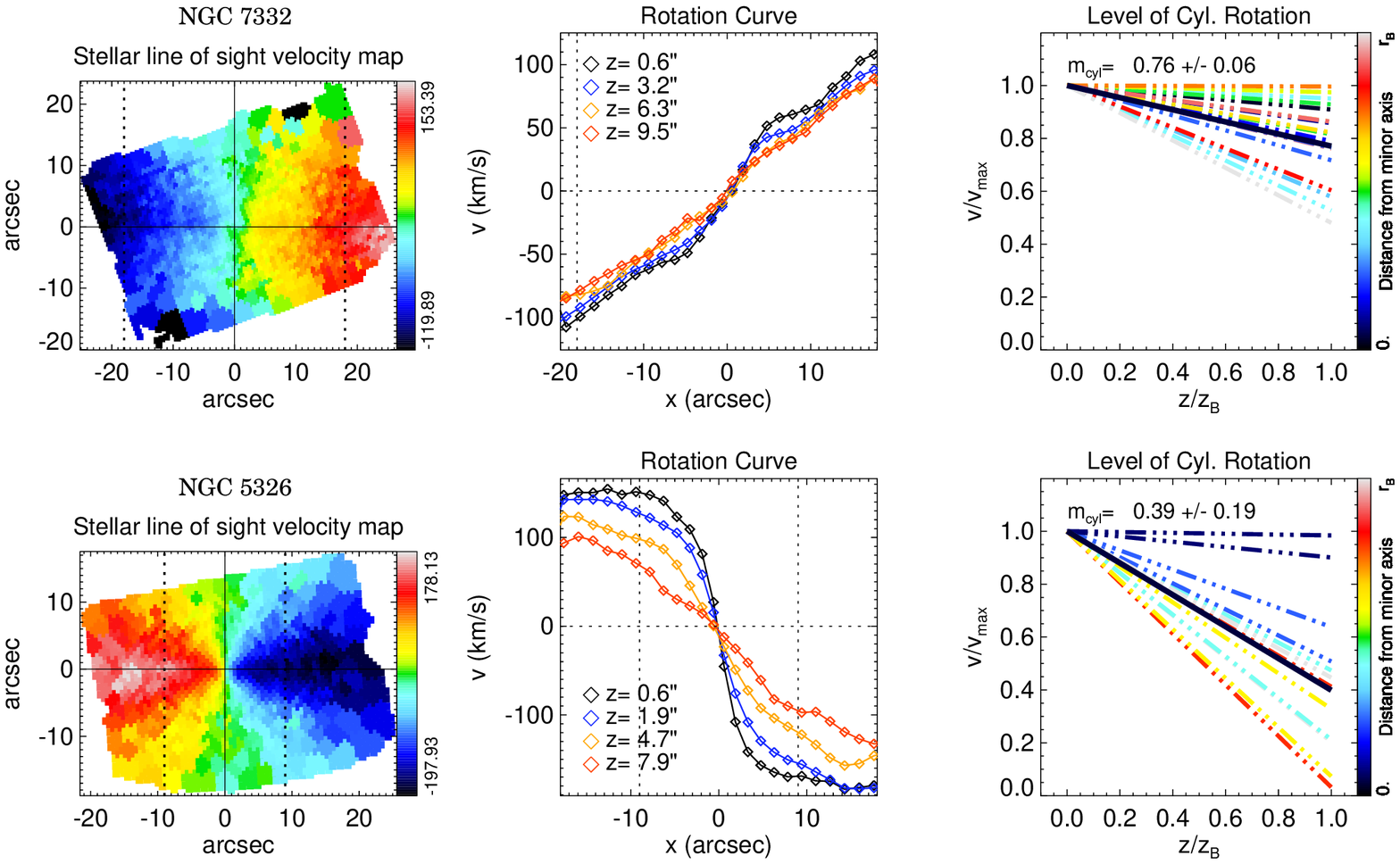}
\caption{(\textit{Left}) The velocity map of NGC\,7332 (a cylindrical rotating
system) and NGC\,5326 (a non cylindrical rotator). In these plots, velocity maps
are rotated so that the horizontal axis is parallel to the major axis of the
galaxies. Vertical dotted lines mark the bulge boundaries. (\textit{Middle}) The
stellar line of sight velocity parallel to the major axis of galaxies at
different height ($z$) from the disk plane. (Right) Line-of-sight velocity
gradient along the minor axis of galaxies at different distance from the minor
axis. Profiles are color-coded according to the distance from the minor axis.
Solid black profile represents the average value. We define \mcyl\,=\,$m_{\rm avg}
+ 1$ as a quantity to express the level of cylindrical rotation, in which $m_{\rm avg}$ is the slope of the solid black line . See text for the definition of the various quantities $m_{cyl}$, $z_{B}$ and $V_{max}$. }
\label{fig:cylrot}
\end{figure*}

\section{Quantifying cylindrical rotation}
\label{sec:cylrot}

\citet{korm1982} were the first who reported in the boxy bulge of NGC\,4565
line-of-sight velocities that remained almost constant with increasing height
above the disk plane. They called this unusual kinematic feature 
\textit{cylindrical rotation}. This feature was later also reported in several
 studies \citep[e.g.][]{bure1999,falc2003a,falc2004,falc2006},
which all confirmed its relation to BP bulges. The good agreement between
numerical models and observed kinematics of barred galaxies suggests that
cylindrical rotation can be considered as a characteristic behaviour of such
systems.

\citet{will2011} used long slit spectroscopy to determine that the degree
of cylindrical rotation varies significantly among BP bulges. This study argued 
that BP bulges do not form a homogeneous class of objects. This
conclusion, however, was based on a small sample of galaxies. Most of the
observed deviations from perfect cylindrical rotation were associated to velocity
profiles in slits relatively close to the disk plane, where fast rotating
components, thick disks or dust could significantly affect the observed
kinematics. It is important to note that the wide range of cylindrical rotation
levels in their sample was established qualitatively by visual inspection.

\citet{saha2013} were the first to introduce a quantitative criterion to measure
deviations from pure cylindrical rotation in edge-on disk galaxies. They
computed the mean mass weighted line-of-sight velocity of stars, within a set of
pseudo-slits along the major axis of model galaxies within the bulge dominated
region. They then plotted these values against the slit locations above the
galactic mid-plane and linearised this profile. They argued that the slope of
this linearised profile can be considered as a quantity to express the level of
cylindrical rotation. We tested this method with our integral-field data and
realised that points farther away from the minor axis have more weight in the
calculations and, consequently, the level of cylindrical rotation becomes very
sensitive to the definition of bulge boundaries. Moreover, the adopted
normalization factor in this method is obtained from the line-of-sight velocity
of regions close to disk plane, where the velocity maps of galaxies are strongly
affected by complicated nature of contributing components. These issues probably arise
by the fact that this method is constructed based on a simplified picture of BP
bulges in their $N-$body models. 

Inspired by the method by \citet{saha2013}, we present here a more robust 
approach to quantify the level of cylindrical rotation. It is optimised to work
with integral-field data. For this purpose, on the clean (i.e., dust-free) side of 
our galaxies, we define a set of pseudo-slits, parallel to
the minor axis of the galaxies at different positions along the major axis, covering
 the whole horizontal extent of the bulge from $-x_{B}$ to $+x_{B}$. 
 The outer slits (with respect to the minor axis) are considered to be wider to compensate for
  the small number of contributed data points (Voronoi-bins), so that the number of 
  voronoi-bins within all slits is almost the same.

For each pseudo-slit, we plot the absolute value of velocities against the absolute
value of height from the disk plane ($z$). We then fit a straight line to this
velocity profile excluding the regions close to disk plane, which are most
likely disrupted by contamination of dust and/or central components
(evaluated by visual inspection of the velocity maps. This region is marked in Figure~\ref{fig:corrprofs}). 

Using this new linear profile, for each pseudo-slit, we evaluate the velocity at $z$=0 (i.e. the disk of the
galaxy). Therefore, we have a set of linearised velocity profiles from
$z$=0 to $z$=$z_{\rm bulge}$ at different distances from the minor axis from $-x_{B}$ to $+x_{B}$. Each profile is normalized to its maximum velocity (i.e. for a given slit placed at $x=x_{i}$ the profile is normalized to $V^{i}_{max} \equiv V^{i}(z=0)$, in which $V$ is the linearized velocity)
 and maximum extent of bulge in the $z$ direction ($z_{\rm B}$).
With this normalization, the slope of this linear profile ranges from -1 to 0.
Finally, we calculate the average slope of the linearised velocity profiles
related to different slits ($m_{\rm avg}$) and define \mcyl\,=\,$m_{\rm avg}
+ 1$ as a quantity to express the level of cylindrical rotation. With this
definition, values of \mcyl\ are generally between 1 which presents pure
cylindrical rotation and 0 where there is no sign of cylindrical rotation.\looseness-2

Figure~\ref{fig:cylrot} illustrates this procedure for two galaxies in our
sample: NGC\,7332, which is a known barred galaxy with a BP bulge and NGC\,5707,
a galaxy with an spherical bulge. The figures clearly shows that for NGC\,7332
stellar rotation changes very little away from the galaxy mid-plane, while
changes in NGC\,5707 are much more pronounced. This is also reflected in the
spread of the linearised vertical profiles shown in red and blue. Similar
figures for the entire sample are presented in Appendix~\ref{app:cylrot}. We
list the measured values of \mcyl\ in Table~\ref{tab:cylrot}.

Figure~\ref{fig:cyl_comp} demonstrates a comparison between the method
introduced by \citet{saha2013} and our new approach to evaluate the level of 
cylindrical rotation in NGC\,0678, a barred galaxy with BP bulge. 
In the top panel, the bulge radius along the minor axis ($z$) is fixed 
at $z_{B}$ and the level of cylindrical rotation for different values of $x_{B}$ 
has been evaluated using the both methods. In the bottom panel, the $x_{B}$ is fixed at a certain value and level of cylindrical rotation is evaluated for different values of $z_{B}$.

While both methods do not show strong dependency to the definition of $z_{B}$ (bulge 
analysis radius along the minor axis), our new approach is more stable against 
variations along the major axis. This example demonstrates that our method is robust agaist inner disk components, as NGC\,0678 has, and works better than Saha's, at least in cases with noisy profiles.

\begin{figure}
\includegraphics[width=0.5\textwidth]{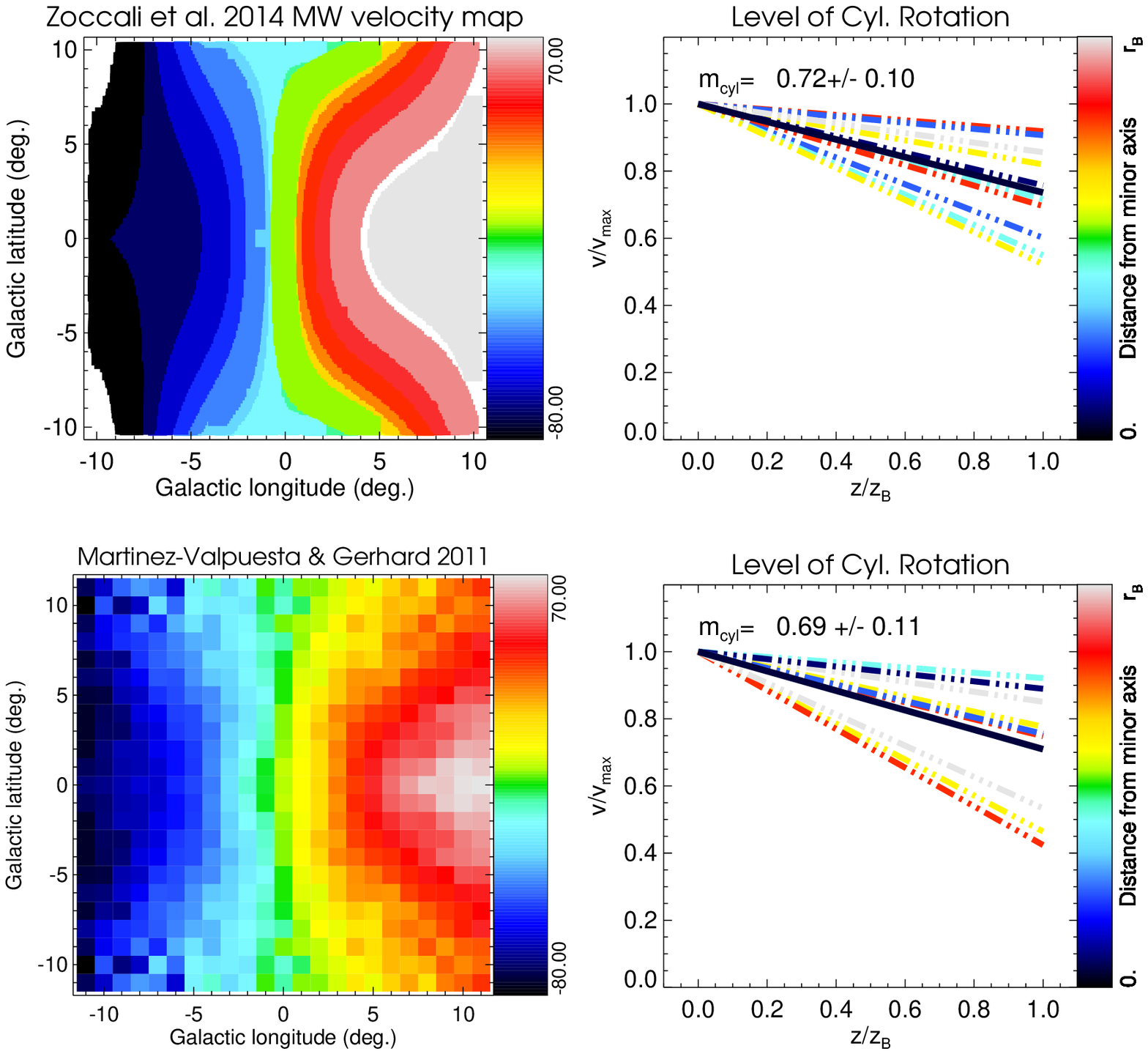}
\caption{(\textit{Top row}) Left panel presents the mean radial velocity map of
the Milky Way bulge in the longitude-latitude plane constructed from the
measured rotation profiles of \citet{zocc2014}. The right panel shows the
linearised velocity gradient along the minor axis of Milky Way at different
distanced from the minor axis. Description of the plot is the same as Figure~\ref{fig:cylrot} (Righ Panel). As discussed in the text, we use this profiles to
evaluate the level of cylindrical rotation (\mcyl). (\textit{Bottom row}) In the
left panel we show a self-consistent $N-$body simulation of the Milky Way bulge
from \citet{,mart2011, gerh2012, mart2013}. The right panel is equivalent to the top, right but
for this simulation. Interestingly, \mcyl\ for this simulation matches fairly
well that coming from real observations.}
\label{fig:mw_cyl}
\end{figure}
\begin{figure}
	\centering
	\includegraphics[width=0.35\textwidth]{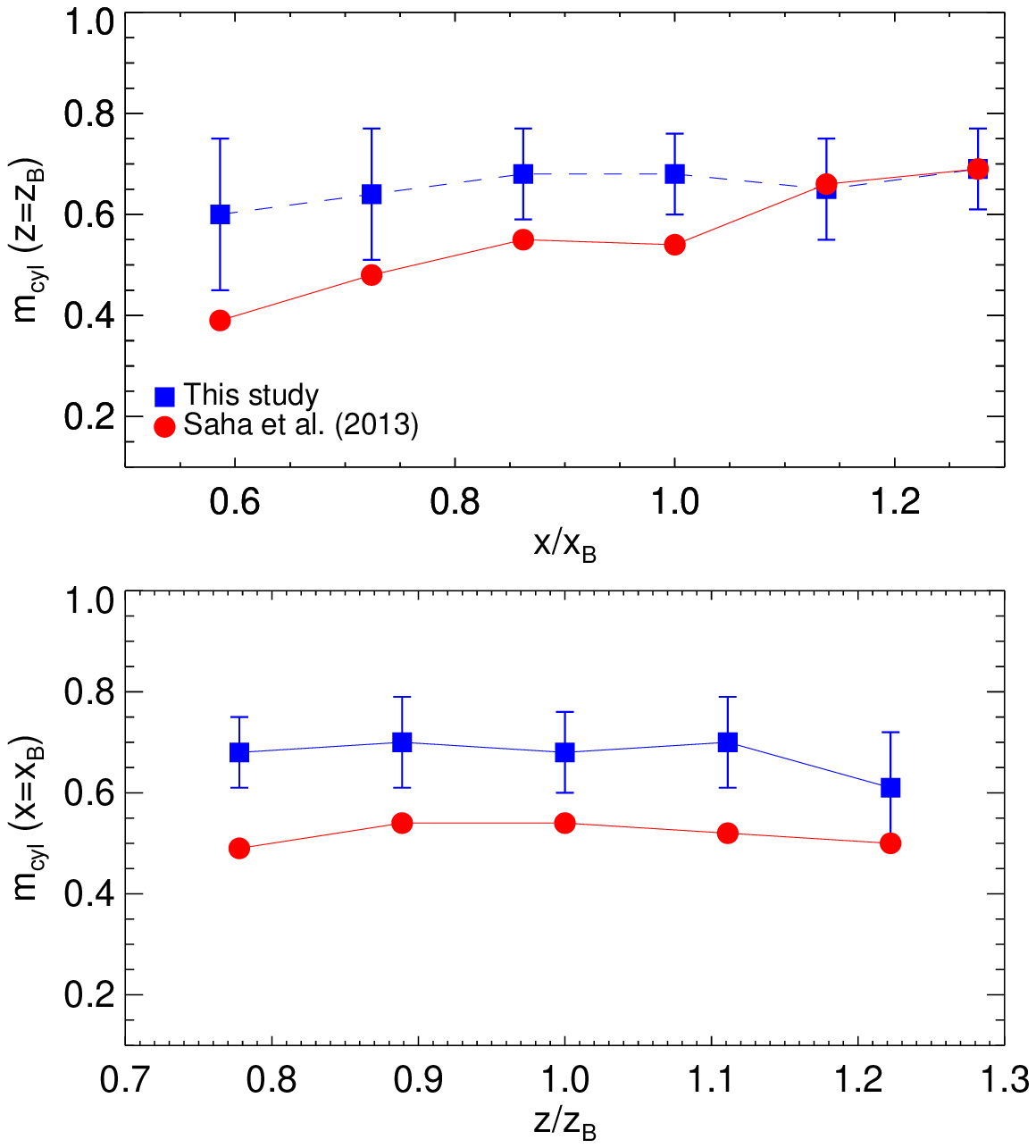}
	\caption{Level of cylindrical rotation at different major axis (top panel) and minor axis (bottom panel) radii of NGC\,0678. In upper panel, the bulge radius along the minor axis ($z$) is fixed to $z_{B}$ and the level of cylindrical rotation for different values of $x_{B}$ has been evaluated using both \citet{saha2013} method and the approach, presented in this study . In lower panel, the $x$ is fixed at a certain value ($x_{B}$) and level of cylindrical rotation is evaluated for different values of $z_{B}$. As this figure shows, the level of cyl rotation based on our proposed approach show less sensitivity to the definition of $x_{B}$ compare to the \citet{saha2013} method. The dependency of both methods to the bulge analysis radius along the minor axis ($x_{B}$) is negligible.}
	\label{fig:cyl_comp}
\end{figure}

\section{Cylindrical rotation in the Milky Way bulge}
\label{sec:mw_cyl}

It is now widely accepted that the Milky Way (MW) is an example of a barred
galaxy with an BP bulge \citep[e.g.][]{blit1991, stan1994, dwek1995, babu2005,
ratt2007}. As we are situated in the disk plane, the detailed analysis of
Galactic bulge kinematics is a non-trivial task. During the past decade,
different surveys of the Galactic bulge have described a clearer picture of the
Galactic bulge. Cylindrical rotation was first established in the MW bulge from
data at latitudes between $b=-2$ and $b=-8$ by the BRAVA survey \citep{rich2007,
howa2009, kund2012}. More recently, the Giraffe Inner Bulge Survey
\citep{zocc2014}, carried out at the ESO-VLT with the multi-fibre spectrograph
FLAMES, targeted around 5000 red clump giants. The primary goal of this survey
is to characterize the kinematics, metallicity distribution, and element ratio
of those stars across 31 fields in the Galactic bulge region. These results
confirmed previous claims of cylindrical rotation. \citet{zocc2014} also
presented radial velocity, and velocity dispersion maps of the MW bulge, by
interpolating the measured values of radial velocity and velocity dispersion at
different observed locations at negative latitudes. They also used data at
$b=4.5$ in order to verify the assumption of symmetry with respect to galactic
plane. In the context of our work in this paper, these maps are very useful as
they allow for direct comparison of the Galactic bulge kinematic maps with
those of external galaxies.

We have taken advantage of this opportunity and applied our method to quantify
the level of cylindrical rotation in the MW bulge. 
Figure.~\ref{fig:mw_cyl} (top row) shows the velocity map extracted from the \citet{zocc2014} article. Based on our measurements, the degree of cylindrical rotation (\mcyl), within the whole available area ($-10<l<+10$ and $-10<b<+10$), evaluated by \citet{zocc2014} is $0.72\pm0.1$.

In addition, we have also applied this method to the same scaled area from the
$N$--\,body simulation of \citet[][hearafter MVG11]{mart2011}, which was known to match remarkably
well the structure of the MW bulge \citep{gerh2012,mart2013}.
In the MVG11 model of the MW, the bar length is $\sim 4.5 kpc$, oriented at an angle $\alpha=25^{\circ}$ with respect to the line from the Galactic center to the observer.
Perhaps not surprisingly, the \mcyl\ value for this simulation is very much consistent with what we measured from the \citet{zocc2014} observations. However, as the color profiles in right panels of figure 5 show, the velocity pattern for these two cases is not alike
\footnote{Note that, the velocity map, presented by \citet{zocc2014} is an infered map from discrete data points, not homegenously covering the $l$-$b$ grid, while in the MVG11 model of MW bulge (Bottom row of Figure~\ref{fig:mw_cyl}), all the particles in the line of sight have been included. This could explain the discrepancies between these two velocity fields.}, while the representative (mean) values, for both cases, show a high level of cylindrical rotation. These results show that the MW bulge is cylindrically rotating at the same level as the strongest objects in our sample of external galaxies.

To address any concern regarding the field size mismatch between the MW and our sample galaxies, for estimating $m_{\rm cyl}$, we refer to Figure~\ref{fig:sim-cyl}, in which we also present the variation of \mcyl\ with inclination and PA based on the simulated galaxies of MVG11 for which the disk subtraction has been carried out as in the sample galaxies. The variation in the region of interest ($Inclination\; i=90^{\circ}$ and $Bar\:PA=65^{\circ}$), which matches that of the MW, is smooth and the \mcyl\ value is quite similar ($\sim 0.7$) to the value obtained from \citet{zocc2014}.

\begin{table*}
\caption {Bar signatures and Bulge classification.}
\label{tab:cylrot}
\begin{center}
\begin{tabular}{lcccccccccccc}
\hline
\hline
\textbf{NGC\,number} & \textbf{0678} & \textbf{5326} & \textbf{5422} & \textbf{5475} & \textbf{5689} & \textbf{5707} & \textbf{5746} & \textbf{5838} & \textbf{5965} & \textbf{6010} & \textbf{7332} & \textbf{7457} \\
\hline
1) Double-hump rotation curve       & Y  & N & Y  & N & Y & N  & Y & Y & Y & N?  & Y & N \\
2) Broad central $\sigma$  peak     & N? & N & N? & N & Y & N  & Y & Y & Y & N  & Y & N \\
with plateau at moderate radii      & ~  & ~ & ~  & ~ & ~ & ~  & ~ & ~ & ~ & ~  & ~ & ~ \\
3) $V-h_3$ correlation over the     & Y  & N & Y  & N & Y & N  & Y & Y & Y & N  & Y & N \\
 projected bar length               & ~  & ~ & ~  & ~ & ~ & ~  & ~ & ~ & ~ & ~  & ~ & ~ \\

\hdashline

Central $V-h_3$ anti-correlation & Y  & N & Y  & Y & Y & N  & Y & Y & Y & Y  & Y & N \\
\hline
Boxy/Peanut Bulge                   & Y  & N & Y  & N & Y & N  & Y & N & Y & N  & Y & N \\
\hline 
\hline                                  
\textbf{Degree of Cyl. Rotation (\mcyl)} & 0.66 & 0.39 & 0.70 & 0.44 & 0.64 & 0.22 & 0.63 & 0.40 & 0.62 & 0.42 & 0.76 & 0.78 \\
&\tiny $\pm0.09$ &\tiny$ \pm0.19$ &\tiny$ \pm0.07$ &\tiny $\pm0.17$ &\tiny$ \pm0.09$  &\tiny$ \pm0.21$ & \tiny $\pm0.08$ & \tiny $\pm0.17$ &\tiny$ \pm0.10$ &\tiny $\pm0.17$ &\tiny$ \pm0.06$ &\tiny $\pm0.08$ \\
\hline
\textbf{Bar}                        & Y  & N & Y  & N & Y & N  & Y & Y & Y & Y? & Y & N \\
\hline
\hline
\end{tabular}
\end{center}
\begin{flushleft}
\small NOTES: In this table, the "Y" refers to Yes and "N" refers to No answer.
The Question mark (?) next to few answers demonstrates the uncertainty in those
answers. The upper section of the table shows the characteristic of the $V$,
$\sigma$ and $h_3$ profiles, which are extracted only on the clean€ (i.e.
dust-free) side of our galaxies, along the major axis at different height from
the disk plane. The (anti)correlation between the $V$ and $h_3$ are studied in
details in \S\ref{sec:corrprofs} and the results are presented in
\S\ref{fig:corrprofs}. The second panel of this table presents our
classification of the bulge-type by visually inspecting the unsharp-masked images as well as the residual images after
subtracting an exponential disk model from the original images (we refer the reader to
\S\ref{sec:sam_pro} for more details). The last panels present the degree or
cylindrical rotation (\mcyl) for our galaxies, as described in \S\ref{sec:cylrot}.
Based on the bar diagnostics, presented, we label our galaxies as barred or
non-barred in the last row.
\end{flushleft}
\end{table*}

\section{Discussion}
\label{sec:discussion}

In this section we summarise our results regarding the presence of bars and
level of cylindrical rotation for the entire sample, and discuss the limitations
of the diagnostics described above. In Table~\ref{tab:cylrot}, we list our
findings in terms of kinematic features, boxiness, and cylindrical rotation of
the bulges in our sample.

\subsection{Kinematic signatures of bars in our sample}
\label{sec:kinfeatures}

We analysed the two dimensional kinematics maps of a sample of 12 intermediate
inclined ($i > 60^{\circ}$) galaxies. As established with our image
decomposition and unsharp masked images (see Figure~\ref{fig:allphot}), 50\% of galaxies in this
sample contain BP bulges (NGC\,0678, NGC\,5422, NGC\,5689, NGC\,5746, NGC\,5965,
NGC\,7332). The linear correlation profiles of all these 6 galaxies reveal
strong $V-{\rm h}_3$ anti-correlation within the central, few arcsec region.
This feature is associated to observed fast rotating, axisymmetric inner disks,
most likely formed secularly through bar-driven processes. Due to the
axisymmetric nature of such structures, the related kinematic signatures will be
largely independent of viewing angle. Another common feature in these 6 galaxies
is the positive correlations between the velocity and $h_3$ beyond the central
regions in areas dominated by the BP bulge light. All these galaxies also
display double-hump velocity profiles and fairly broad velocity dispersion
profiles (see Appendix~\ref{app:cylrot}). In addition they present the highest
levels of cylindrical rotation. All this information combined confirm
that the observed BP bulges are strongly linked to bar structures. 

At the opposite end, there are 4 galaxies (NGC\,5326, NGC\,5475, NGC\,5707,
NGC\,6010) with nearly spherical bulges and no indication of bars being present
based on their kinematic information. None of these galaxies are known in the
literature to host bars. The level of cylindrical rotation in these bulges are
also the lowest of the sample and thus confirms the kinematic diagnostics.
NGC\,6010 is a special case among these four galaxies. The level of positive
$V-{\rm h}_3$ correlation above the bulge dominated region is quite large, it also shows a strong anti-correlation in the inner parts (possibly indicating
the presence of an inner disk). 
There is, however, no sign of strong double-hump velocity profiles or
broad velocity dispersion major axis profiles.

As discussed in \S\ref{sec:corrprofs}, we suggest that NGC\,6010 
	is a barred galaxies with an spherically-symmetric central component, in which 
	the vertical extend of the bar is beyond the classical bulge in the center. 
	With this configuration we can interpret the unusual kinematical behavior of this galaxy which 
	is very similar to unbarred systems within the bulge zone and simillar to barred 
	galaxies beyond that.


As already pointed out in \S\ref{sec:corrprofs}, another interesting case is
NGC\,5838. It has a fairly round bulge but presents kinematic features
remarkably similar to those of strongly barred galaxies listed above. It shows a
strong anti-correlation in the inner parts and positive correlation away from
the disk plane, in the bulge dominated regions. The major axis kinematic
profiles also show clear double-hump rotation and broad central velocity
dispersion. The only feature that differs from those exhibited by strong barred
systems is the low level of cylindrical rotation measured. As we explain in the
next section, this is expected if the bar is oriented parallel to our
line-of-sight. Therefore we propose that this galaxy is in fact barred.

There is only one exception in our sample (NGC\,7457) in which the level of
cylindrical rotation is considerably high (\mcyl\,$>$\,0.7), but the
photometric and kinematics properties of this galaxy suggest there is no bar
present. \citet{silc2002} analysed the two-dimensional kinematics maps of this
peculiar galaxy in detail and proposed that this galaxy is the result of
two superposed quasi-counter-rotating stellar disks with different inclinations.
At this point, it is not obvious whether the high level of cylindrical rotation
measured could be linked to this peculiar orbital configuration or not, although
some studies suggest that axisymmetric configurations can in certain circumstances
also give rise to cylindrical rotation \cite[e.g.][]{rowl1988}. NGC\,7457 is
therefore a very peculiar case in our study.

Our results suggest that 8 out of 12 galaxies in our sample contain bars, with
at least one of them (NGC\,5838), possibly 2 (NGC\,6010), being new
identifications. Based on the results presented here we confirm that the major-axis stellar 
kinematic diagnostics presented by \citet{bure2005} and the extension of this method to the 2D line-of-sight 
kinematics of simulated disk galaxies by \citet{2015MNRAS.450.2514I} are reliable discriminantors for determination of 
the presence of bars in highly inclined systems. We show that the
$V-{\rm h}_3$ correlation profiles are particularly useful for this purpose.
Cylindrical rotation is another important feature of barred systems, but it is not always
a good discriminant as it depends on the galaxy's inclination and bar
orientation along the line-of-sight. In the following section we characterise
these changes in cylindrical rotation using $N$--\,body simulations.

\subsection{The effect of bar orientation and galaxy inclination on the level of cylindrical rotation}

\begin{figure}
\centering
\includegraphics[width=0.45\textwidth]{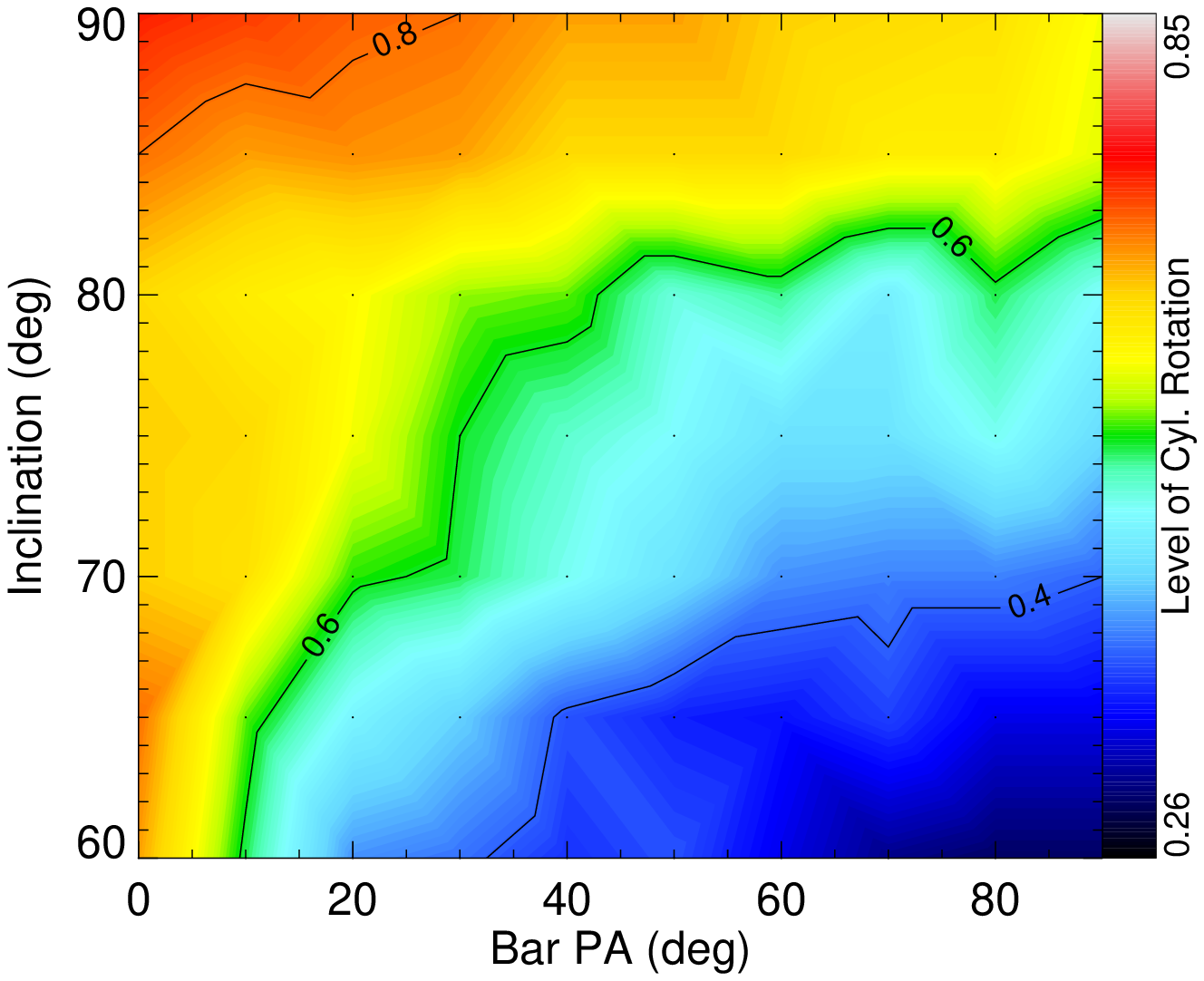}
\caption{Level of cylindrical rotation (\mcyl) in a $N$--\,body simulation of a
bar-unstable disk for different galaxy inclinations and bar position angle. The
range of galaxy inclinations explored here is similar to the ones of our sample
of galaxies. In this figure, a side-on bar (i.e. perpendicular to the
line-of-sight) has a Bar PA of $0^{\circ}$ , while an end-on bar (i.e. parallel
to the line-of-sight) has $90^{o}$. Small dots mark the reference locations of
\mcyl\ values in our simulations. The colour map shows intermediate levels,
interpolated from the reference points.}
\label{fig:sim-cyl}
\end{figure}

Our observational results confirm a close link between the morphology and
kinematics of barred disk galaxies. We also show that cylindrical rotation
appears to be the sensitive to the particular configuration between the
galaxy inclination and the bar position angle with respect to our line-of-sight.
To illustrate this dependency we use kinematic maps extracted from
self-consistent three-dimensional $N$--\,body simulations of a boxy bulge and bar which formed from a bar-unstable disk by \citet[][hereafter IMV06]{mart2006} and \citet[][hereafter MVG11]{mart2011}.
We treat these maps in the same way we
deal with the observed data, considering the method to evaluate the bulge analysis window, thus computed the \mcyl\ values in a consistent
manner.
The I1 series of the MVG11 that we have used here, evolved from an initially exponential disk with $Q = 1.5$ embedded in a dark matter halo, and developed a prominent boxy bulge through a buckling instability after $\sim1.5\;Gyr$. Figure~\ref{fig:sim-cyl} presents the result of this exercise for a particular snapshot in this simulation, when the bar and its BP bulge has fully developed.
The figure shows how the level of cylindrical rotation changes as a function of
galaxy inclination and bar position angle. As shown in \citet{mart2008}, the
strength of the bar correlates with the strength of the BP bulge, and therefore
the absolute values displayed in the figure depend somewhat on the strength of
the bar. Nevertheless, the general behavior is the same for all cases studied.
The figure shows that when the bar is oriented perpendicular to our
line-of-sight, cylindrical rotation hardly changes with galaxy inclination (for
values above 60$^\circ$ at least). This suggests that, while the presence of the
main disk of the galaxy is becoming more and more important along our line-of-sight as
we depart from edge-on inclination, its effect is not significant when it comes
to affect the \mcyl\ measurements. At edge-on inclination, it seems that the
orientation of the bar is not an obstacle to measure high levels of cylindrical
rotation in barred systems. Nevertheless, the situation changes as inclination
decreases and the bar is oriented parallel to our line-of-sight. We believe this
effect may be at the root of the low levels of cylindrical rotation observed in
NGC\,5838, while the other kinematic diagnostics clearly suggest the presence of
a bar in the galaxy.

\subsection{Cylindrical rotation in bulges of nearby galaxies}

In previous sections we have established a new methodology to compute the level
of cylindrical rotation (\mcyl) in galaxies and also described some of the
potential shortcomings of this parameter. Here we will concentrate on presenting
the distribution of \mcyl\ values for our sample of galaxies, and also what one
would expect in general based on results from our numerical simulations.

Figure~\ref{fig:cyldist} (top panel) shows the distribution of \mcyl\ values for
our sample. Perhaps not surprisingly, all the galaxies with BP bulges display
levels of cylindrical rotation above $\approx$0.6. As mentioned earlier, the
MW bulge \mcyl\ value is very consistent with those of BP bulges. Galaxies
with no signs to host bars, on the other hand, have \mcyl\ values below
$\approx$0.5. The only peculiar case of a non-barred galaxy with high \mcyl\
values is NGC\,7457 (see \S\ref{sec:kinfeatures} for details). At the low \mcyl\
end we also have the two galaxies that we believe harbour bars in an
end-on orientation (NGC\,5838, NGC\,6010).\looseness-2

While our sample is admittedly limited, it serves to see the range of possible
values of \mcyl\ in galaxies. The location of BP and spherical bulges is clearly
distinct, but with the low number statistics is difficult to assess the degree
of overlap between these two families of objects. 

In order to understand the expected true distribution of this parameter, we have resorted to the set of $N$--\,body simulations (IMV06, MVG11) used in the previous section, complemented with a set of numerical simulations of barred galaxies with classical bulges, evolved from an exponential disk and a classical bulge embedded in a live dark matter halo, with $B/D=0.25$ and $Q=1.5$. For these simulations, "non- barred" refers to the stage when the bar is flat and "barred", after the boxy bulge has formed and the bar has regrown. The results are shown in Figure~\ref{fig:cyldist} (bottom panel). We present the results for non-barred galaxies in blue and barred galaxies in red.

For barred systems, we plot values of \mcyl\ for a wide range of different
time snapshots and combinations of galaxy inclination (above 60$^\circ$) and bar position angle. The distributions of the two kind of objects clearly
separated with mean values $0.35\pm0.10$ for non barred systems and $0.65\pm0.15$ for barred galaxies.
The distribution for barred galaxies, however, shows cases with
low \mcyl\ values, overlapping regions dominated by galaxies with no bars. These
are cases where the bar position angle and galaxy inclination are in
unfavourable orientations (see Figure~\ref{fig:sim-cyl}). 
This result is in agreement with previous studies by \cite{atha2002} and \citet{2015MNRAS.450.2514I} which indicate the effect of the viewing angles on the measured 
level of cylindrical rotation in simulated disk galaxies. They indicate a tendency for cylindrical rotation to weaken 
when going from side-on to end-on views. They also stress that the importance of this effect does 
not follow a clear trend with BP strength and link such behaviour to the different orbital structure in 
each bar and on various properties of the periodic orbits of the main families.

Our observations and the complementary numerical tests thus confirm that
cylindrical rotation is a reliable indicator for the presence of bars only when
its levels are high. Low \mcyl\ values are not definite prove of the absence of
bars.

\begin{figure}
\centering
\includegraphics[width=0.5\textwidth]{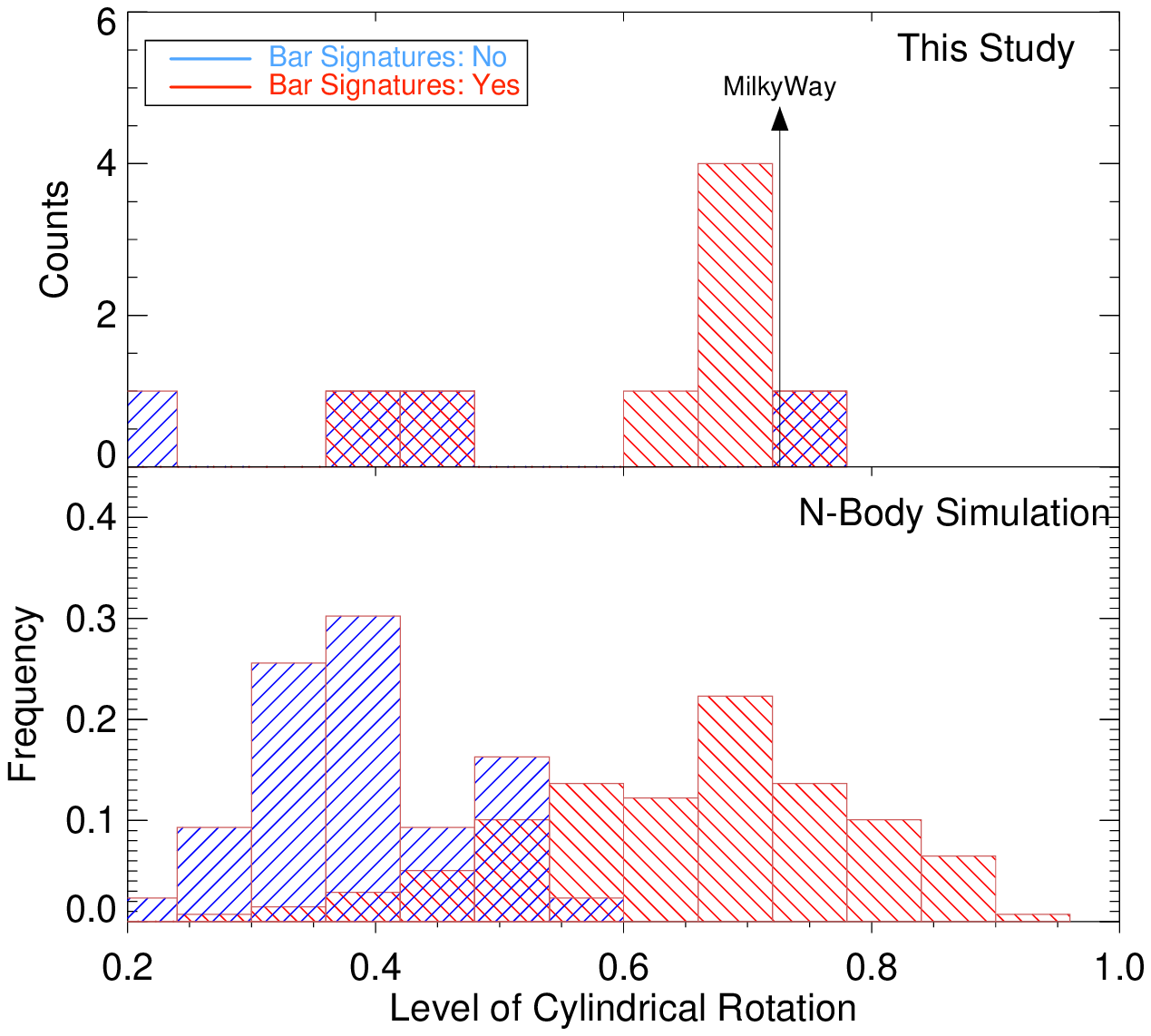}
\caption{Distribution of the level of cylindrical rotation (\mcyl) for
observations (top panel) and in $N$--\,body simulations (bottom panel). For the
observations, galaxies with clear signatures of bars are marked in red, while
those with no signs are in blue. In the simulations, non-barred galaxies are
marked in blue and barred systems in red.}
\label{fig:cyldist}
\end{figure}

\section{Conclusions}
\label{sec:conclusions}

This paper investigates the connection between bulge morphology and kinematics
in 12 mid to high-inclination ($i > 60^{\circ}$) disk galaxies, from the
\citet{balc1994} sample, observed with the \sauron\ integral-field spectrograph.
The goal was to unveil the presence of hidden bars, using the major axis kinematic
diagnostics developed by \citep{bure2005} and \citep[][optimised for 2D kinematic data]{2015MNRAS.450.2514I} and establish the overall level of
cylindrical rotation (\mcyl) in those bulges. For the latter, we developed a
new method (inspired by the work of \citealt{saha2013}) that quantifies the
importance of this property in our bulges.

We find that the strong positive correlation between the stellar velocity ($V$)
and the $h_3$ Gauss-Hermite parameter appears to be the most reliable indicator
for the presence of bars among all other bar diagnostics. This is true even in
situations when the bar orientation is parallel to line-of-sight and
consequently the bar not clearly visible in photometric data. We exploited this
feature and determined that 8 galaxies in our sample harbour bar structures.
While some of them were known or were very obvious cases, we have identified 3
new cases based on their kinematic properties (NGC\,5422, NGC\,5838, NGC\,6010).

Bulges in our sample display a wide range of cylindrical rotation values, with
BP bulges showing the largest levels (\mcyl\,$\approx$\,0.7). Using the same
methodology, we have established that the Milky Way bulge shows comparable
levels of cylindrical rotation (\mcyl\,=\,0.72\,$\pm$\,0.1) to the BP bulges of
external galaxies. Nearly spherical bulges (e.g. non-barred galaxies) display
values around \mcyl\,$\approx$\,0.35.

Our study with $N$--\,body simulations confirms that high levels of cylindrical
rotation can be considered in general as a common feature of strongly barred
galaxies (e.g. BP bulges). The opposite is not true though, as particular
orientations of the bar along the line-of-sight (e.g. end-on bars) display
relatively low values of \mcyl (levels in fact comparable to unbarred galaxies).
Therefore cylindrical rotation cannot be considered a reliable kinematic
property to identify the full population of barred galaxies. The same
simulations suggest that the distribution of \mcyl\ for barred and non-barred
galaxies have clearly distinct mean values. Nonetheless, there is significant
overlap in the low-end caused by non-favourable galaxy and barred orientations
(i.e. galaxy low-inclination, and bar position angle close to 90$^\circ$) or
flat bar phases.

In summary, focusing on the study of the kinematic properties of intermediate
inclined galaxies, we have unveiled the presence of bars and also
established the level of cylindrical rotation in bulges in general. In the
following paper of this series we will relate these properties to the stellar
content in these bulges, paying particular attention to the impact of
cylindrical rotation to the age and metallicity gradients away from the
mid-plane of these galaxies.

\section*{Acknowledgments}
The first author wish to thank the School of Astronomy, IPM for providing support
 while working on this paper. A.M. also acknowledges the Isaac Newton Group of
  Telescopes (ING) and the Instituto de Astrof\'isica de Canarias (IAC) for
hospitality and support while this paper was in progress. We also thank an anonymous referee for a lot of constructive comments. The authors acknowledge support from the Spanish Ministry of Economy and 
Competitiveness (MINECO) through grants AYA2009-11137, AYA2013-48226-C3-1-P,
 AYA2013-46886-P and AYA2014-58308-P. J.F.B also acknowledges financial support 
 from the DAGAL network from the People Programme
(Marie Curie Actions) of the European Unions Seventh Framework Programme
FP7/2007-2013/ under REA grant agreement number PITN-GA-2011-289313.

Funding for SDSS-III has been provided by the Alfred P. Sloan Foundation, the
Participating Institutions, the National Science Foundation, and the U.S.
Department of Energy Office of Science. The SDSS-III web site is
http://www.sdss3.org/. SDSS-III is managed by the Astrophysical Research
Consortium for the Participating Institutions of the SDSS-III Collaboration
including the University of Arizona, the Brazilian Participation Group,
Brookhaven National Laboratory, Carnegie Mellon University, University of
Florida, the French Participation Group, the German Participation Group, Harvard
University, the Instituto de Astrof\'isica de Canarias, the Michigan State/Notre
Dame/JINA Participation Group, Johns Hopkins University, Lawrence Berkeley
National Laboratory, Max Planck Institute for Astrophysics, Max Planck Institute
for Extraterrestrial Physics, New Mexico State University, New York University,
Ohio State University, Pennsylvania State University, University of Portsmouth,
Princeton University, the Spanish Participation Group, University of Tokyo,
University of Utah, Vanderbilt University, University of Virginia, University of
Washington, and Yale University. 

\bibliographystyle{mn2e}
\bibliography{mybib}

\label{lastpage}

\appendix

\section{Kinematic maps for individual galaxies}
\label{app:cylrot}

For each galaxy we present here the
integrated flux, stellar velocity, $\sigma$, $h_3$ and $h_4$ maps. The bulge
radius along the major axis is marked with black dash lines. We also present
radial profiles for the velocity, $\sigma$, $h_3$ and $h_4$ along the major axis
of galaxies at 4 different heights from the mid-plane. These heights ($z$) are
selected so that, one profile is close to disk plane and one is close to the
bulge border in $z$ direction and two profiles in between. We also show the
degree of cylindrical rotation \mcyl\ for each galaxy.

 \begin{figure*}
 	\captionsetup[subfigure]{labelformat=empty}
 	\centering
 	\begin{subfigure}{0.5\textwidth}
 		\centering
 		\caption{\Large{\textbf{NGC\,0678}}}
 		\includegraphics[width=1.0\textwidth]{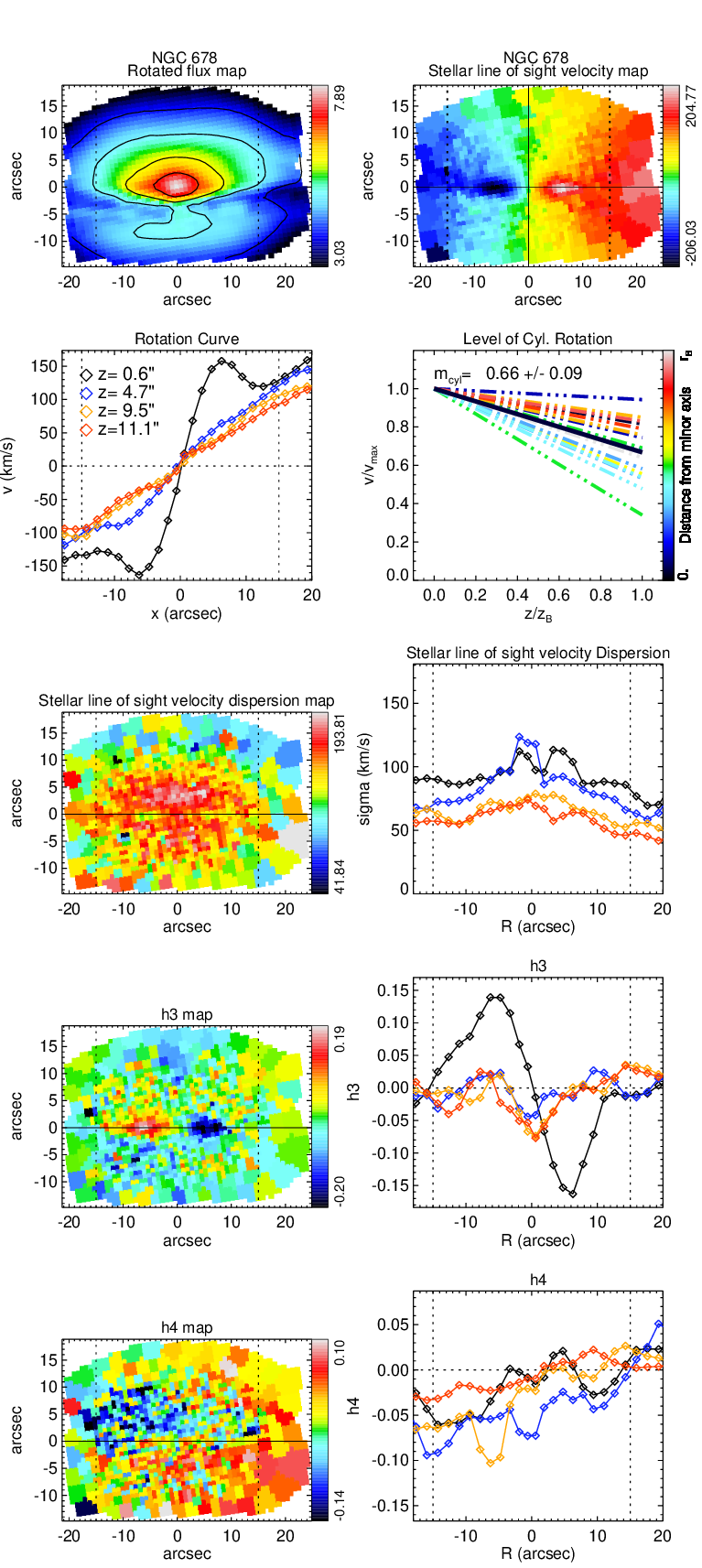}
 		\label{fig:sub1}
 	\end{subfigure}%
 	\begin{subfigure}{0.5\textwidth}
 		\centering
 		\caption{\Large{\textbf{NGC\,5326}}}
 		\includegraphics[width=1.0\textwidth]{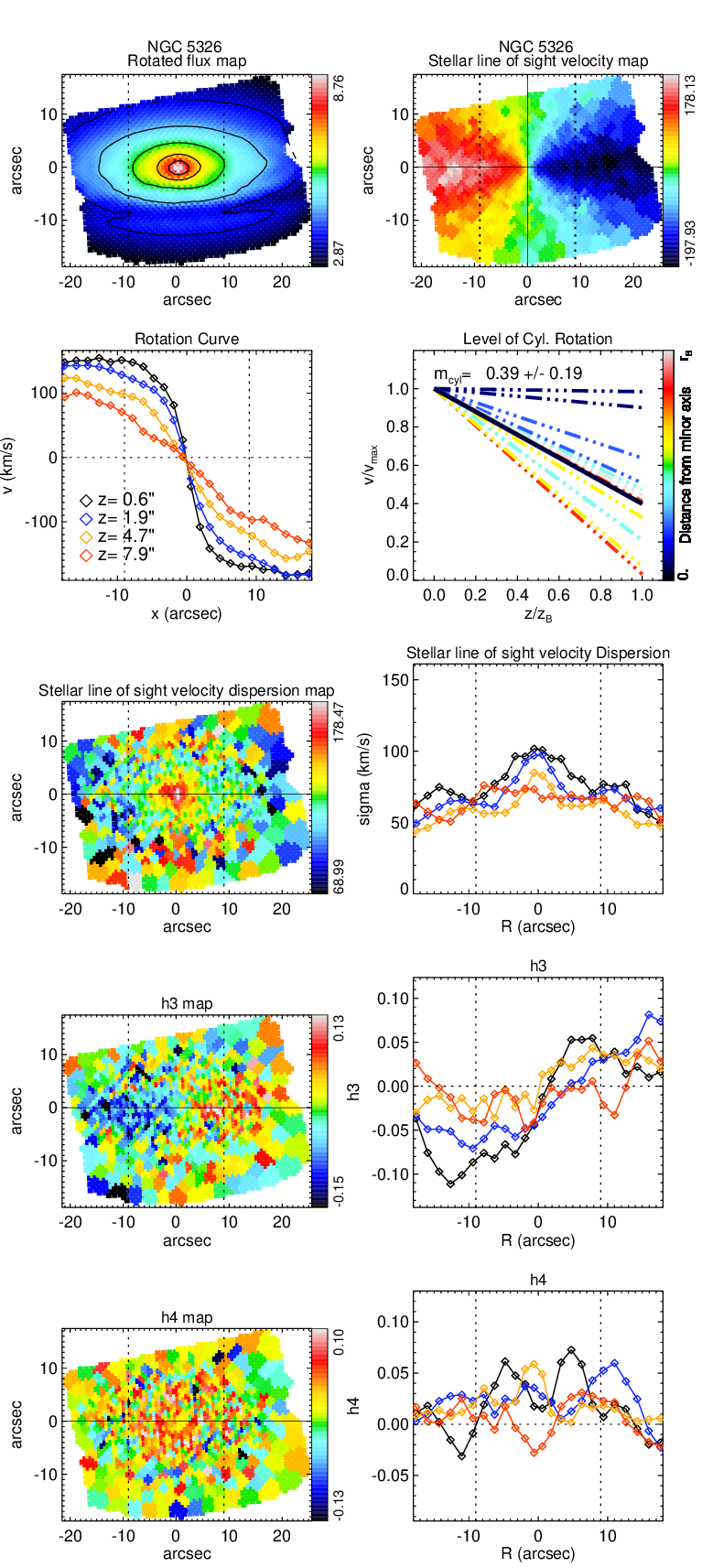}
 		\label{fig:sub2}
 	\end{subfigure}
 	\setcounter{figure}{1}
 	\centering
 	\centering
 	\caption{Sauron stellar kinematics maps, related profiles along cuts parallel to the major axis at four representative different heights from the disk plane and degree of cyl. rotation $(m_{cyl})$ plot. The vertical dashed lines show the bulge boundary in $r$ direction. Note that, all maps are rotated so that the for each galaxy the x axis are along the major axis of galaxy.}
 \end{figure*}

 \begin{figure*}
 	\captionsetup[subfigure]{labelformat=empty}
 	\centering
 	\begin{subfigure}{0.5\textwidth}
 		\centering
 		\caption{\Large{\textbf{NGC\,5422}}}
 		\includegraphics[width=1.0\textwidth]{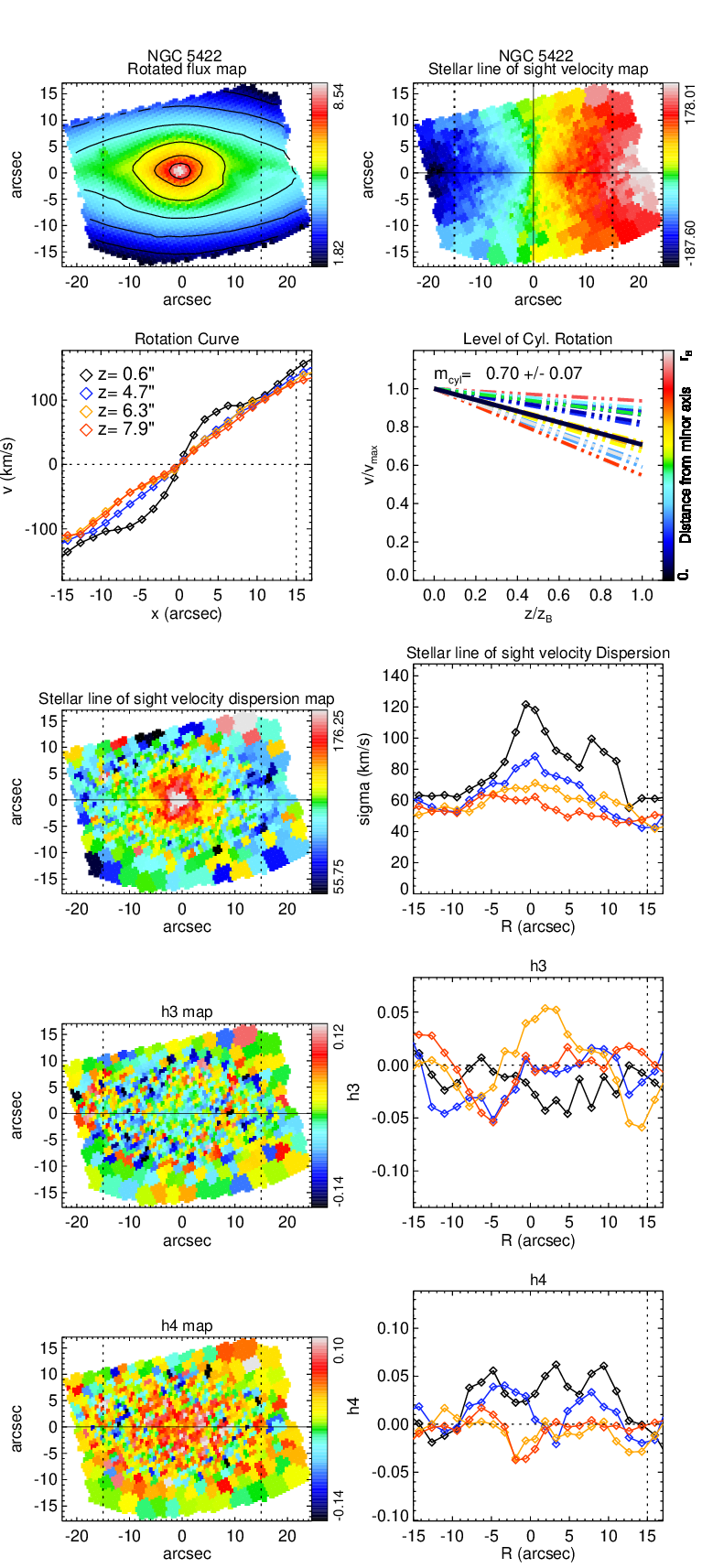}
 	\end{subfigure}%
 	\begin{subfigure}{0.5\textwidth}
 		\centering
 		\caption{\Large{\textbf{NGC\,5475}}}
 		\includegraphics[width=1.0\textwidth]{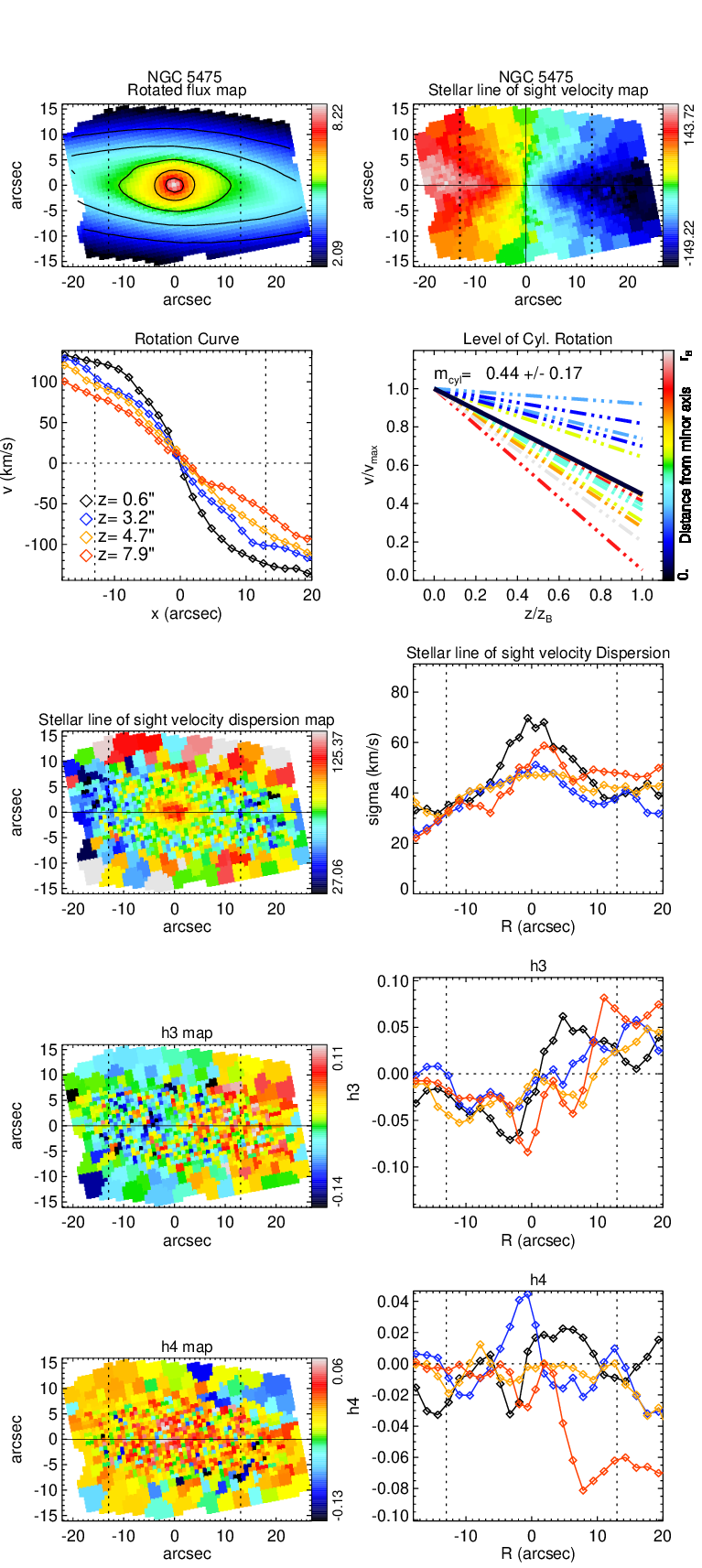}
 	\end{subfigure}
 	\centering
 	\centering
 	\setcounter{figure}{1}
 	\caption{continued}
 \end{figure*}

 \begin{figure*}
 	\captionsetup[subfigure]{labelformat=empty}
 	\centering
 	\begin{subfigure}{0.5\textwidth}
 		\centering
 		\caption{\Large{\textbf{NGC\,5689}}}
 		\includegraphics[width=1.0\textwidth]{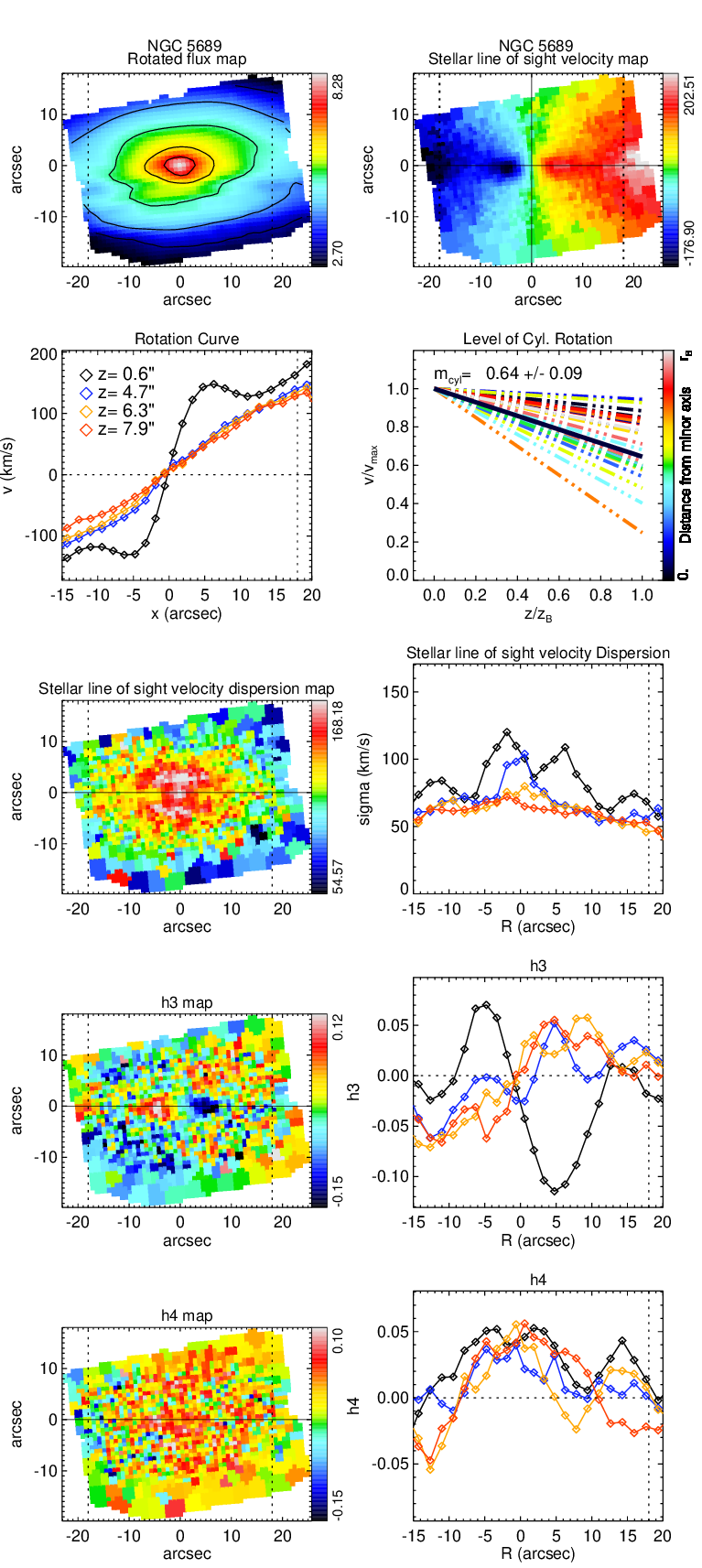}
 	\end{subfigure}%
 	\begin{subfigure}{0.5\textwidth}
 		\centering
 		\caption{\Large{\textbf{NGC\,5707}}}
 		\includegraphics[width=1.0\textwidth]{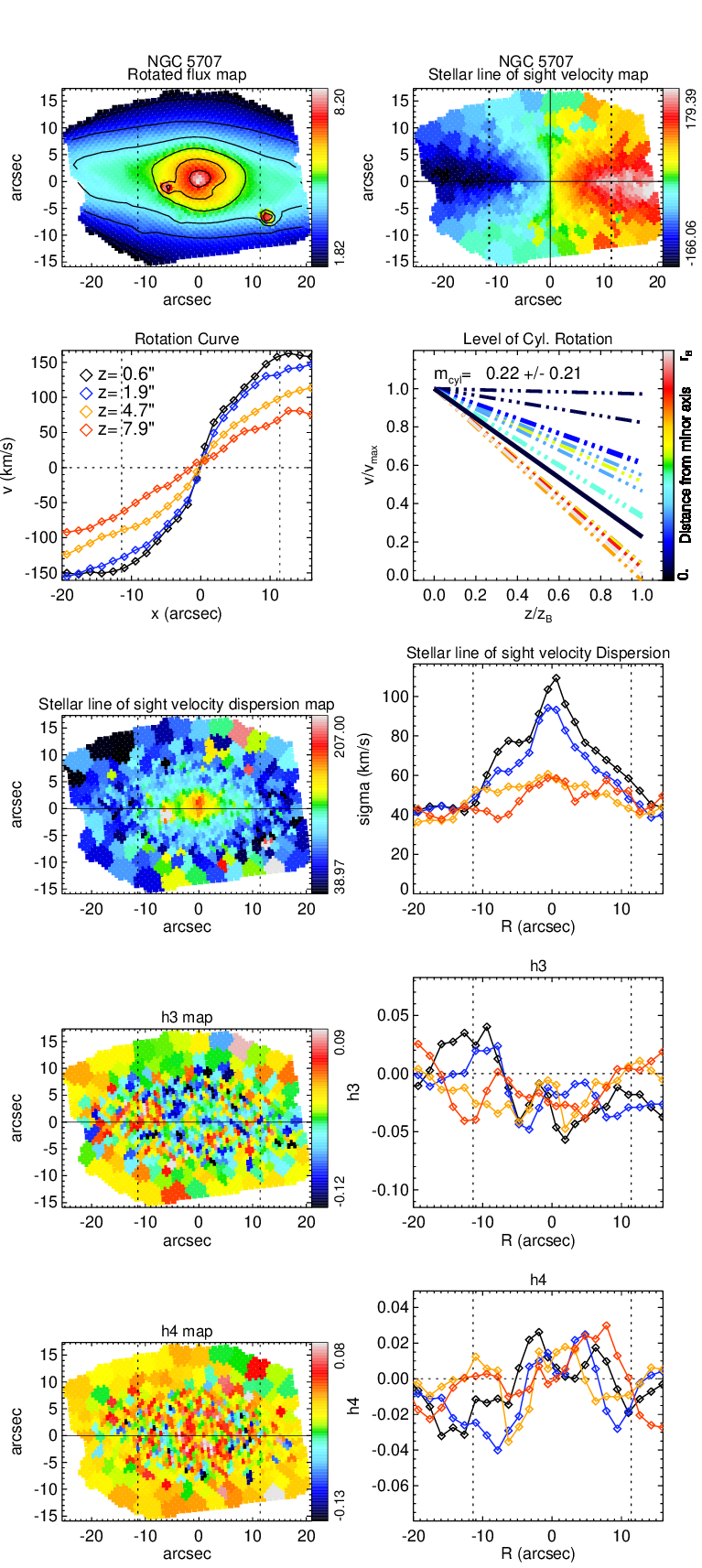}
 	\end{subfigure}
 	\centering
 	\centering
 	\setcounter{figure}{1}
 	\caption{continued}
 \end{figure*}

 \setcounter{figure}{0}
 \begin{figure*}
 	\captionsetup[subfigure]{labelformat=empty}
 	\centering
 	\begin{subfigure}{0.5\textwidth}
 		\centering
 		\caption{\Large{\textbf{NGC\,5746}}}
 		\includegraphics[width=1.0\textwidth]{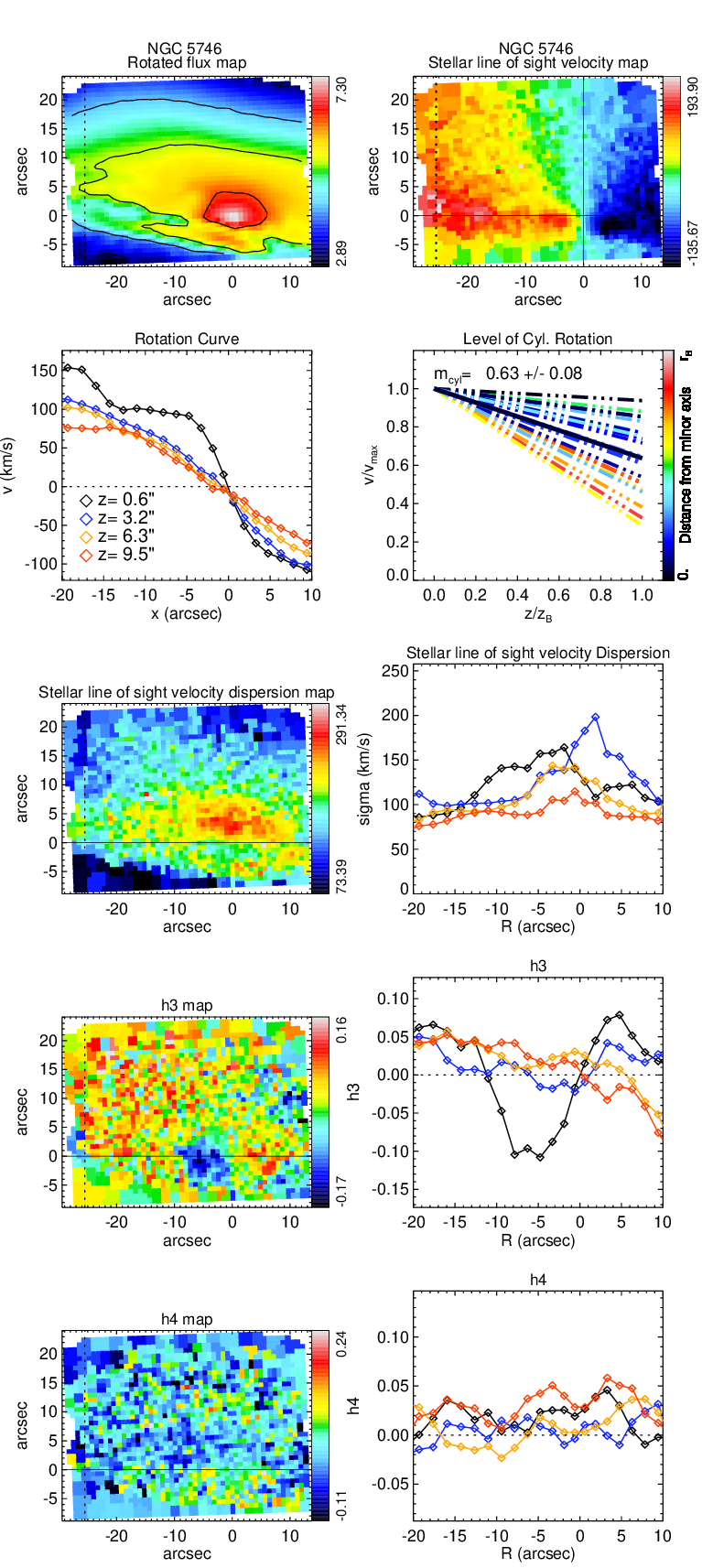}
 	\end{subfigure}%
 	\begin{subfigure}{0.5\textwidth}
 		\centering
 		\caption{\Large{\textbf{NGC\,5838}}}
 		\includegraphics[width=1.0\textwidth]{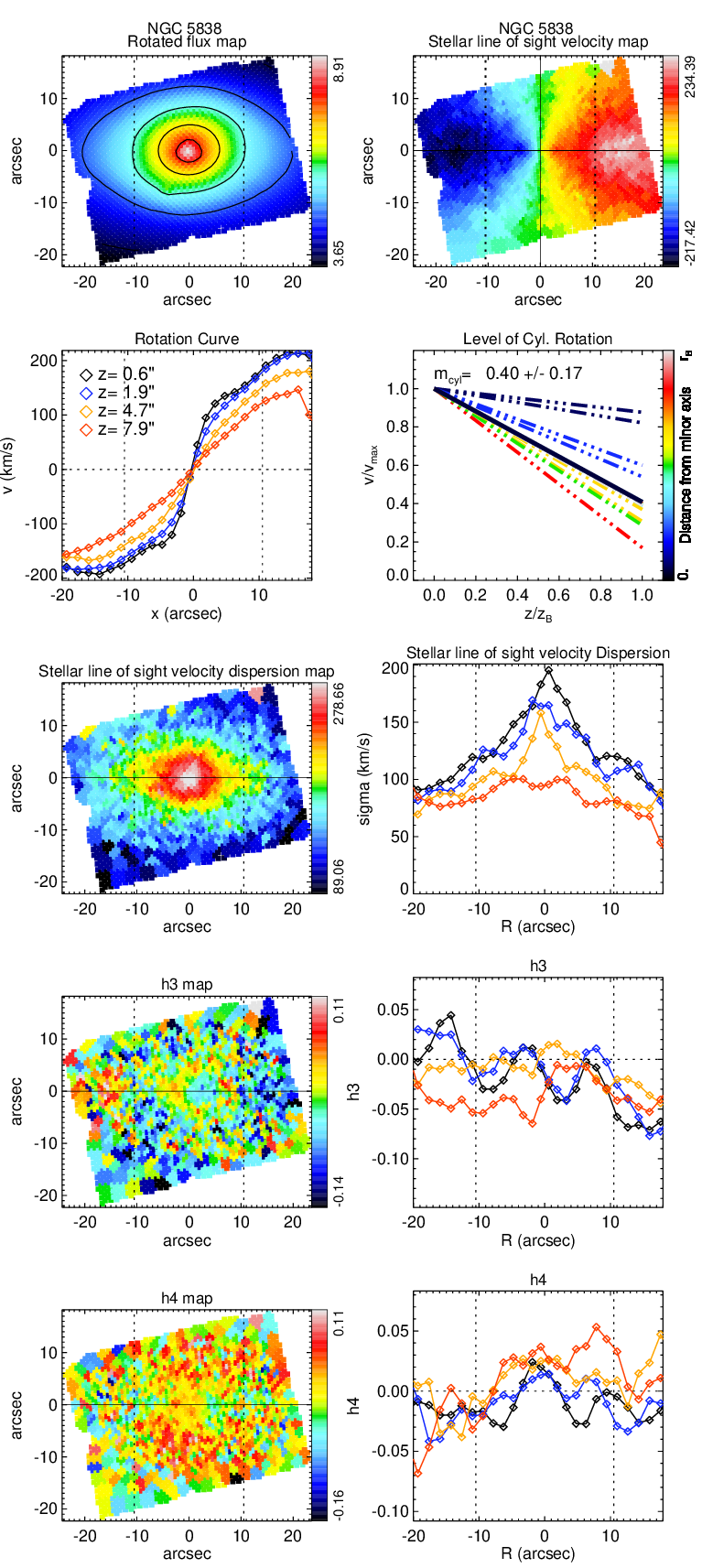}
 	\end{subfigure}
 	\centering
 	\centering
 	\setcounter{figure}{1}
 	\caption{continued}
 \end{figure*}

 \setcounter{figure}{0}
 \begin{figure*}
 	\captionsetup[subfigure]{labelformat=empty}
 	\centering
 	\begin{subfigure}{0.5\textwidth}
 		\centering
 		\caption{\Large{\textbf{NGC\,5965}}}
 		\includegraphics[width=1.0\textwidth]{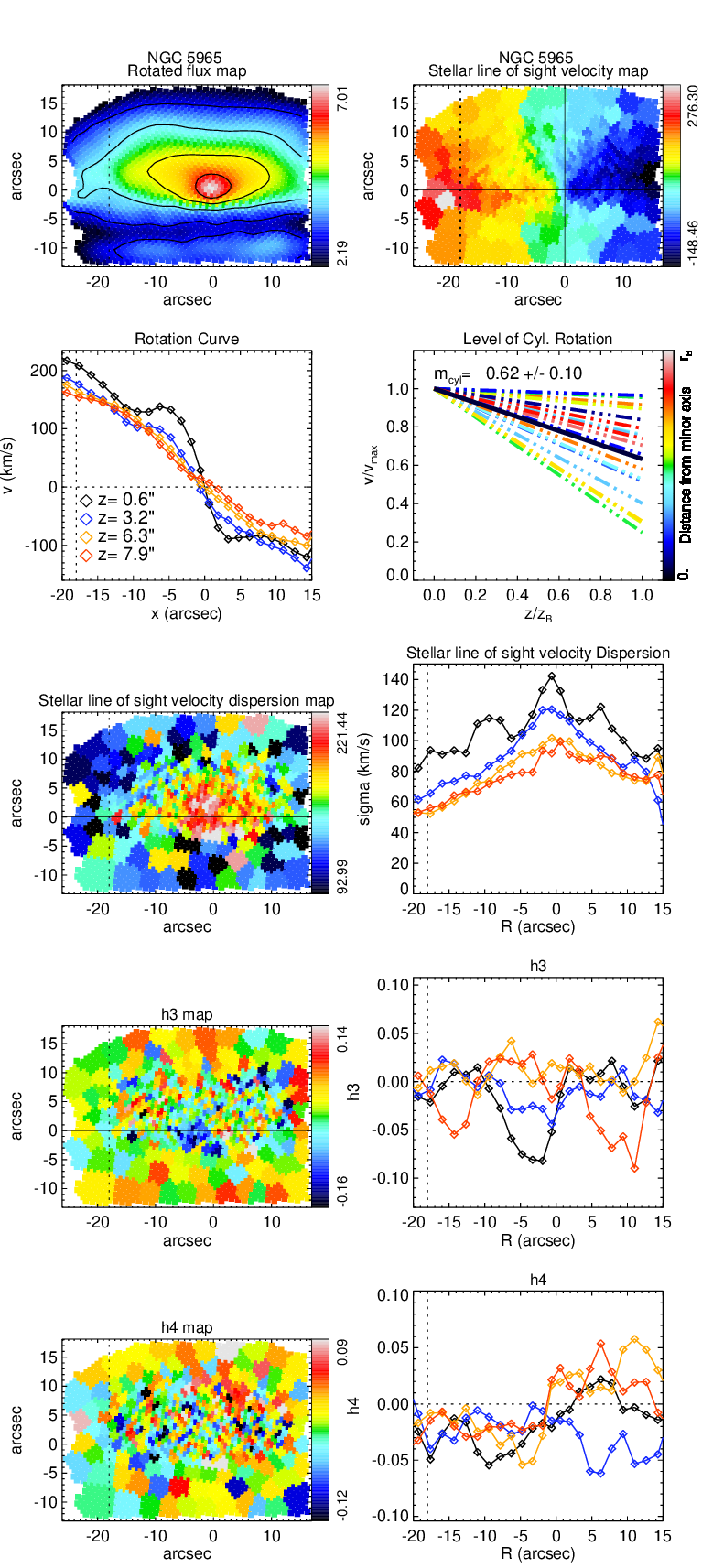}
 	\end{subfigure}%
 	\begin{subfigure}{0.5\textwidth}
 		\centering
 		\caption{\Large{\textbf{NGC\,6010}}}
 		\includegraphics[width=1.0\textwidth]{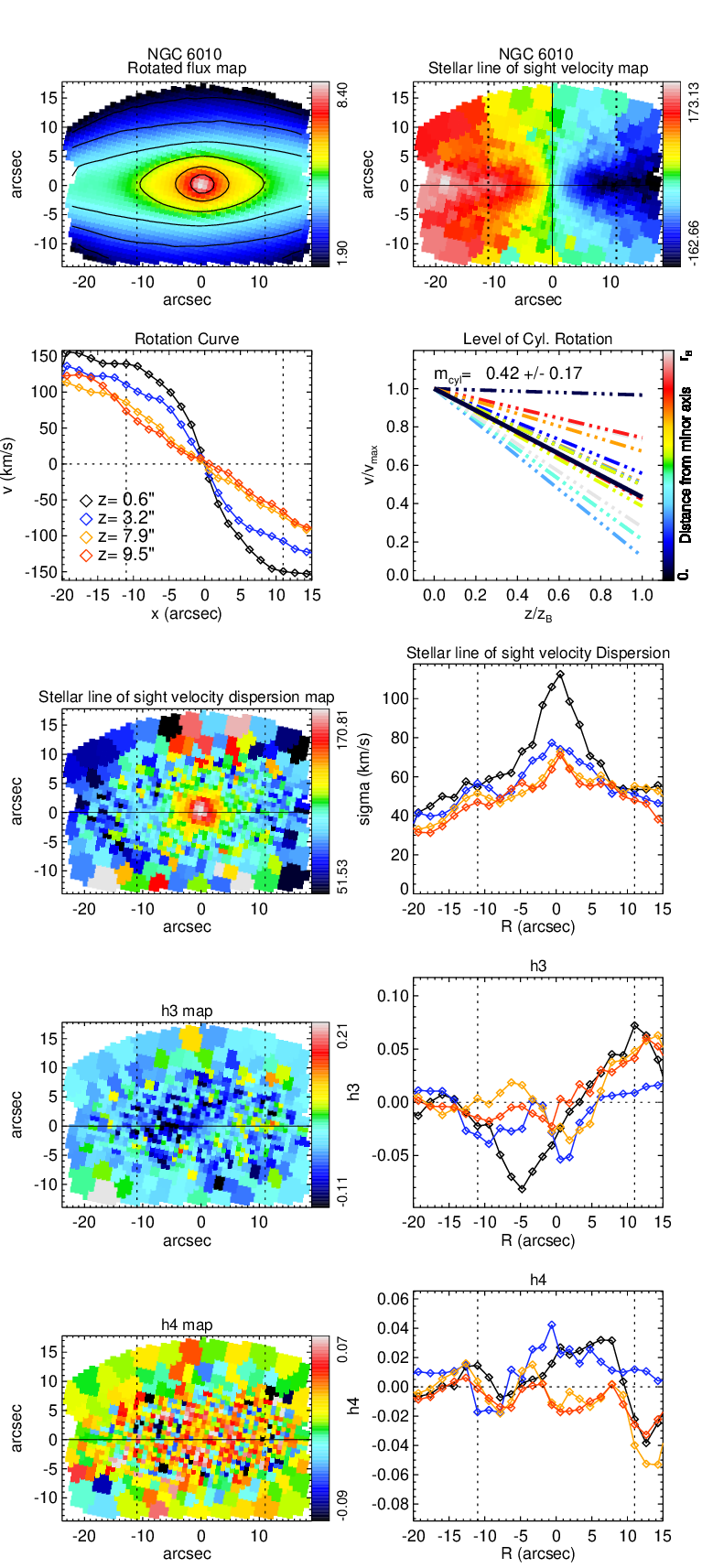}
 	\end{subfigure}
 	\centering
 	\centering
 	\setcounter{figure}{1}
 	\caption{continued}
 \end{figure*}

 \setcounter{figure}{0}
 \begin{figure*}
 	\captionsetup[subfigure]{labelformat=empty}
 	\centering
 	\begin{subfigure}{0.5\textwidth}
 		\centering
 		\caption{\Large{\textbf{NGC\,7332}}}
 		\includegraphics[width=1.0\textwidth]{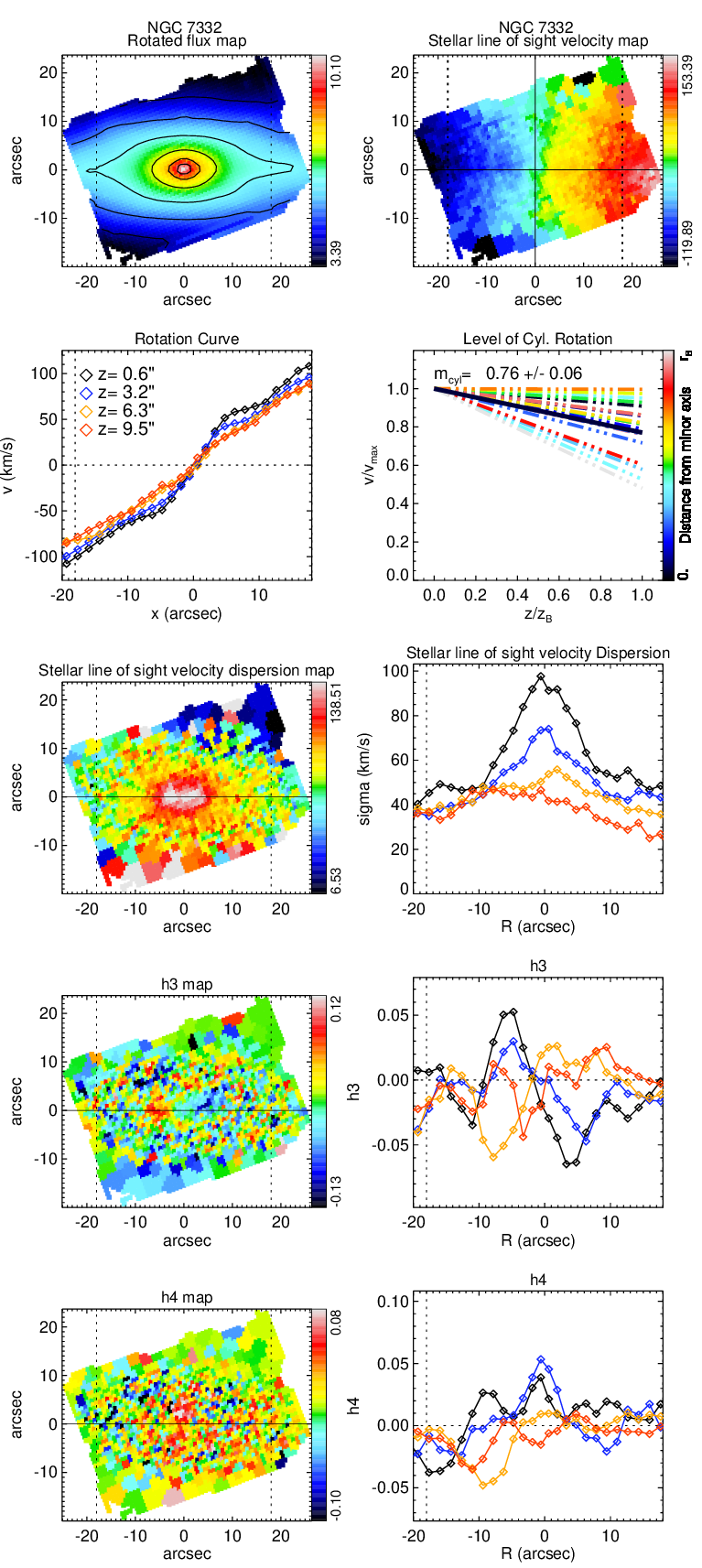}
 	\end{subfigure}%
 	\begin{subfigure}{0.5\textwidth}
 		\centering
 		\caption{\Large{\textbf{NGC\,7457}}}
 		\includegraphics[width=1.0\textwidth]{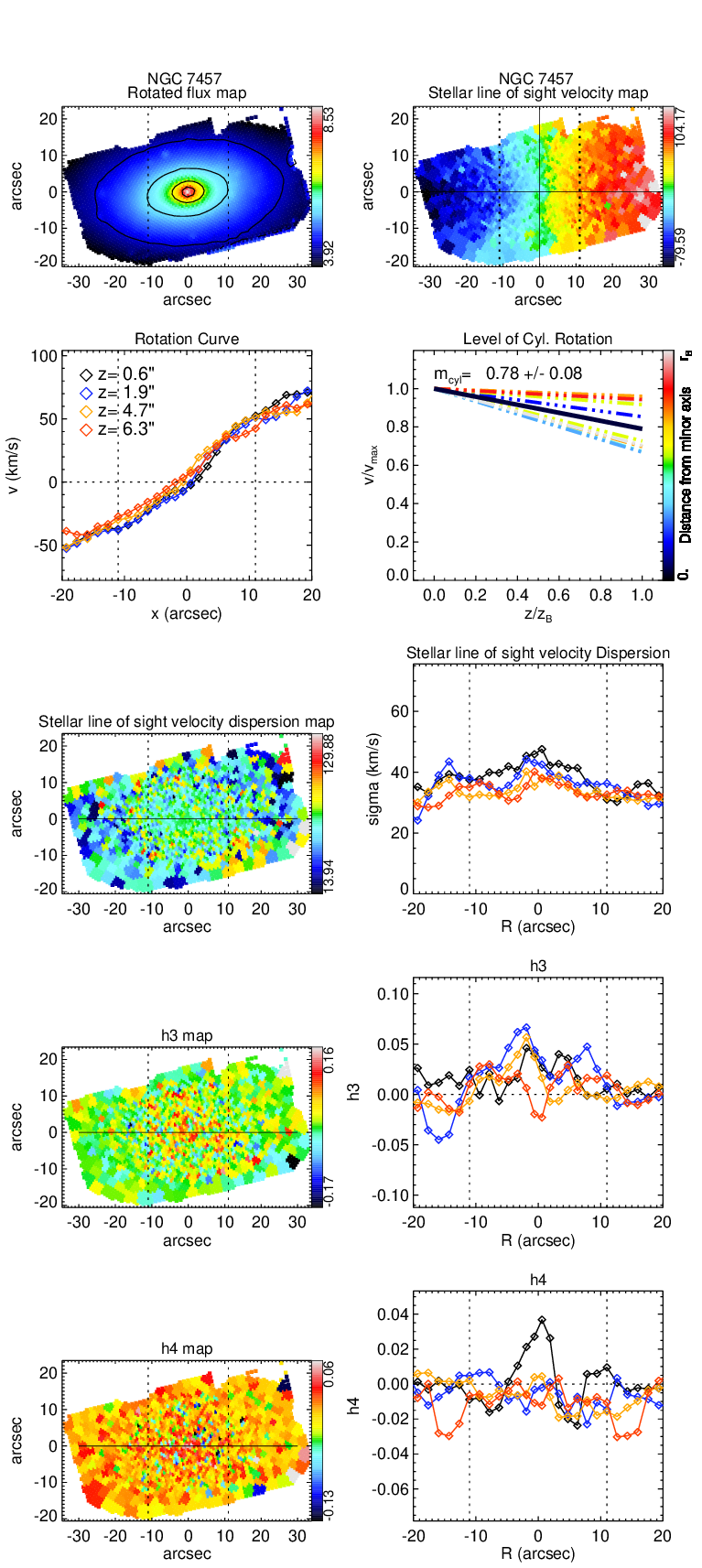}
 	\end{subfigure}
 	\centering
 	\centering
 	\setcounter{figure}{1}
 	\caption{continued}
 \end{figure*}

\end{document}